\documentclass{aa}  

\usepackage{graphicx}
\usepackage{txfonts}
\usepackage{lipsum}

\usepackage{lscape}             
\usepackage{placeins}           
                                
\begin{document}

   \title{Signatures of rocky debris accretion in AF-Type planet hosts}
   \titlerunning{Signatures of rocky debris accretion in F- and A-Type planet hosts}


%

   \author{J. Maldonado\inst{1}\fnmsep\thanks{Corresponding author: jesus.maldonado@inaf.it}
        \and I. Mendigut\'ia\inst{2}
	\and S. Barcel\'o Forteza\inst{3}
	\and G. M. Mirouh \inst{4}
	\and B. Montesinos \inst{2}
	\and E. Villaver \inst{5}
        }

   \institute{INAF - Osservatorio astronomico di Palermo, Piazza del Parlamento 1, 90134 Palermo, Italy
   \and Centro de Astrobiolog\'ia (CAB) CSIC-INTA, ESA-ESAC Campus, 28692, Villanueva de la Ca\~{n}ada, Madrid, Spain
   \and Institute of Applied Computing \& Community Code (IAC$^3$). University of the Balearic Islands, E-07122, Palma de Mallorca, Spain
   \and Departamento de F\'isica Te\'orica y del Cosmos, Universidad de Granada, Campus de Fuentenueva s/n, 18071, Granada, Spain 
   \and Instituto de Astrof\'isica de Canarias, 38200 La Laguna, Tenerife, Spain
  }

   \date{Received September 30, 2026}

 
  \abstract
   {The gas-giant planet-metallicity correlation plays a fundamental role in our understanding of planet formation.
   However, that correlation is mostly derived from observations of solar-type stars.}
   {We aim to test whether the correlation between the metallicity and chemical abundances of the star and the presence of giant planets
   for solar-type stars still holds for early-type (spectral types A and early-F) stars.} 
   {We determine in a homogeneous way the metallicity and individual abundances of a large sample of A and early-F stars, with
   and without known planetary companions, and discuss their abundance distribution and trends. Our methodology is based on the
   analysis of high-resolution \'echelle spectra. It includes the calculation of the fundamental stellar parameters by fitting the
   profiles of the Balmer lines, as well as individual abundances of different elements such as C, O, Mg, Si, Ca, Ti, Fe, Ni and
   Ba by a multi-parameter fitting procedure of selected spectral regions.}
   {We find that stars with planets generally exhibit higher abundances of both iron-peak and volatile elements compared
   to stars without known planets although in most cases this tendency is not statistically significant.
   The $\rm {C/O}$ ratio remains similar between both samples,
   while the  $\rm {Mg/Si}$ ratio is shifted towards higher values in the general planet host sample.
   When the sample is limited to fully radiative stars (spectral type earlier than F5), the statistical significance of
   carbon overabundances vanishes. In contrast, elements such as  O and Ni remain significantly overabundant.
   The  $\rm {C/O}$  and $\rm {Mg/Si}$ ratios in this radiative subsample are statistically indistinguishable from that of the comparison sample.}
   {The persistent overabundance of  O and Ni, combined with $\rm {C/O}$ and $\rm {Mg/Si}$ ratios similar to those of stars without planets in fully radiative hosts, points to the ongoing surface pollution by volatile-poor, silicate-rich rocky debris, rather than the accretion of gas giants. These findings  might position early-type main-sequence stars as a crucial dynamical missing link in the chemical life-cycle of planetary systems: bridging the initial dust-trapping phase in pre-main sequence transition discs, the dilution of this superficial pollution during the fully convective red giant phase, and the eventual dynamically-driven accretion of asteroid material observed in polluted white dwarfs.}


   \maketitle
   \nolinenumbers

\section{Introduction}\label{introduction}

Spectroscopic analyses of the stellar atmospheres of planet host stars
have been fundamental to set observational constrains on planet
formation theories.
In particular, it has been well-established that the frequency of gas-giant planets is a
strong function of the stellar metallicity
\citep[e.g.][]{1997MNRAS.285..403G,2004A&A...415.1153S,2005ApJ...622.1102F},
which is usually interpreted in the framework of core-accretion models \citep[e.g.][]{1996Icar..124...62P}.
The so-called gas-giant planet metallicity correlation 
was originally established for gas-giant planets orbiting around main-sequence (MS), FGK stars.
Since then, 
many works have tried to prove whether or not it also holds for 
other type of stars 
like red giants \citep[e.g.][]{2013A&A...554A..84M,2022A&A...661A..63W}
or M dwarfs \citep[e.g.][]{2013A&A...551A..36N,2020A&A...644A..68M},
as well as other types of substellar objects like rocky planets
\citep[e.g.][]{2012Natur.486..375B,2014Natur.509..593B}
or
brown dwarfs \citep[e.g.][]{2014MNRAS.439.2781M,2017A&A...602A..38M}.

The emerging picture is one in which different planet formation mechanisms may operate altogether and their relative efficiency might change with the mass of the substellar object, stellar (disc) mass, the formation location within the disc, as well as the chemical enrichment of the protoplanetary disc.
\citet[and references therein]{2019A&A...624A..94M} noticed a breakpoint in the metallicity distribution of planet hosts 
suggesting that the  core-accretion formation mechanism for planet formation achieves its maximum efficiency for planets with masses in the range 0.2–2 M$_{\rm Jup}$, although the results by \citet{2019Geosc...9..105A} show that massive planets might be still formed through different channels depending on the mass (protoplanetary disc) of the host star.
Along these lines, \citet{2025AJ....170..334N}
showed the central role played by the chemical enrichment of the protoplanetary disc by finding 
that the most massive planets are likely to be found around the most metal-rich stars.

Besides iron, volatile elements are important for the chemistry of protoplanetary discs and planets.
The role of elements like Si and Mg on planet formation has also been studied \citep{2006ApJ...643..484R,2009MNRAS.399L.103G,2026ApJ...998..301G},
showing that an enhancement in $\alpha$ elements might facilitate the formation of planets around metal-poor stars.
Overall, planet-hosting stars show similar enrichment histories of refractory elements or 
modest overabundances with respect to stars without known planets \citep[e.g.][]{2003A&A...404..715B,2006A&A...449..723G}. 
Detailed chemical abundance studies have revealed different trends in the abundance versus condensation temperature between samples of planet and non-planet hosts,
often manifesting as a deficit of refractory elements in the Sun and other solar analogues.
Whether these trends are related to the presence of planets or not has been strongly debated
\citep[e.g.][]{2009ApJ...704L..66M,2010A&A...521A..33R,2013A&A...552A...6G,2014A&A...564L..15A,2015A&A...579A..20M,2016A&A...588A..98M}.

The formation and evolution of planetary systems does also depend on the mass of the host star.
However, the impact of stellar mass is still poorly known, as most planets have been discovered around
late-type (later than F8) stars.
Previous chemical studies of early-type stars with planets have focused on the scenario proposed by \cite{2015A&A...582L..10K}
in which planet formation leads to an accretion of metal-poor material on the surface of the star \citep[see also][]{2023A&A...671A.140G}.
More specifically, \cite{2021A&A...647A..49S,2022A&A...668A.157S}
explore the chemical pattern of a sample of early-type stars with planets, searching for a possible
relation between the presence of giant planets, metallic-lined A stars, and the $\lambda$ Bo\"otis chemical pattern.
They found observational support that giant planets orbiting pre-MS stars block the dust of the disc resulting in
a deficit of metallic lines that characterises the
$\lambda$ Bo\"otis-phenomenon. 
Interestingly, this pattern is not observed in mature stars hosting hot Jupiters, which rules out the possibility that these
planets are contributing with metal-poor material through their planetary winds.
Interestingly, they show that the incidence of metallic A stars hosting hot brown dwarfs tends to be higher than
the frequency of metallic A stars in general, 
suggesting that the presence of hot brown dwarfs could play a role in the development of metallic A stars.

More recently, \cite{2025A&A...695A..27M} revisited the planet-metallicity correlation in a sample
of intermediate-mass stars covering the full evolutionary sequence from the pre-main sequence to the red giant phase.
The lack of a well-established planet-metallicity correlation in pre-MS and MS intermediate-mass stars
is explained by the fact that intermediate-mass stars are mainly radiative and thus the metallicity of the star does not reflect its bulk composition but the composition of the accreted material. When the star leaves the MS and develops a sizeable convective envelope, a strong-planet metallicity correlation is recovered in line with core-accretion
models of planet formation.

Besides the previous works, a detailed analysis of the chemical properties of early-type stars with planets
is lacking in the literature. This is precisely the goal of this paper, in which we compare the 
chemical abundances of a large sample of A and early-F type planet hosts with the abundances measured
in a similar sample of stars but without known planetary companions. 
This paper is organised as follows. 
Sect.~\ref{observations} describes the stellar samples analysed in this work and how stellar parameters
and chemical abundances are obtained.
The abundance distributions are presented in Sect.~\ref{analysis}.
The results are discussed at length in Sect.~\ref{discussion}. 
Lastly, our conclusions follow in Sect.~\ref{conclusions}.

\section{Data and spectroscopic analysis}\label{observations}
\subsection{Stellar sample}

\begin{figure}[htb]
\centering
\includegraphics[scale=0.55]{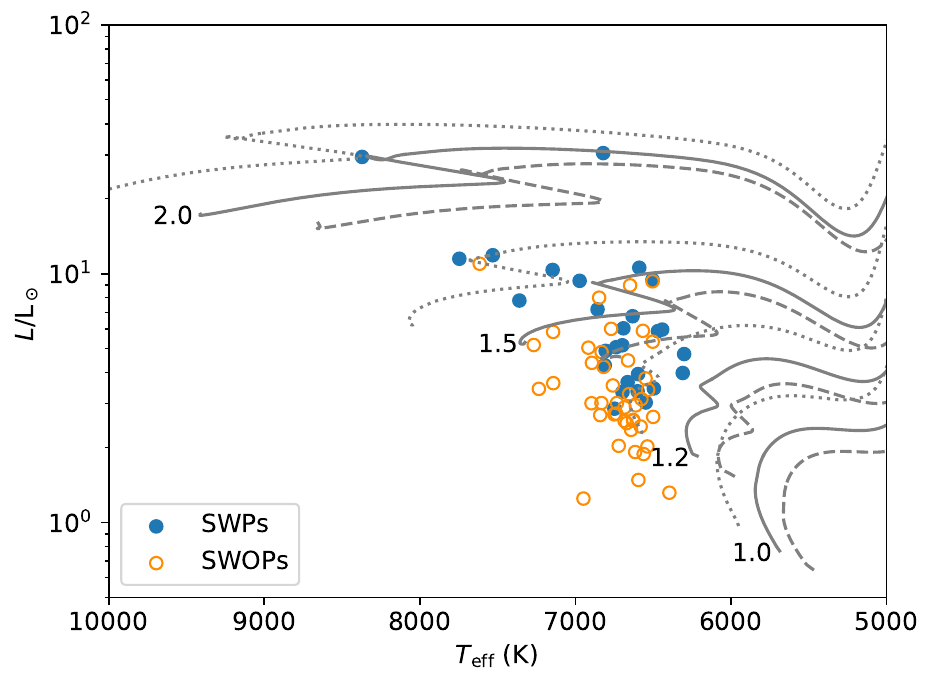}
\caption{
Luminosity versus T$_{\rm eff}$ diagram for the observed stars.
Planet hosts are plotted with filled symbols.
Evolutionary tracks covering the mass range 1.0 - 2.0 M$_{\odot}$
computed with MESA as in \citet[][Sect.~2.2]{2023MNRAS.524.3978M} are overplotted.
Different line styles correspond to
Z = 0.0071 (dotted lines), Z = 0.0142 (solid lines) and Z = 0.0284 (dashed lines).
Error bars are within the symbol size. Temperatures and luminosities are computed
as described in Sect.~\ref{spec_analysis}.
}
\label{hrdiagram}
\end{figure}

A sample of A and early-F 
hosting planets (hereafter stars with planets, SWPs) was built by carefully checking the data
available at the Extrasolar Planets Encyclopaedia\footnote{\url{https://exoplanet.eu/home/}}
\citep{2011A&A...532A..79S} as well as in the NASA Exoplanet Archive\footnote{\url{https://exoplanetarchive.ipac.caltech.edu/}}.
The planets orbiting around these stars
are mostly transiting, with radii ranging between 0.4 and 2.1 Jupiter's radius
and short periods, although some of them (those confirmed by direct imaging) might have periods as large as roughly 60 years.  
In order to compare the properties of the planet hosts with those
of a similar sample of stars without known planets, a subset of early-F and A type stars
was selected from \cite{2009A&A...495..335L}. These stars were
monitored with the HARPS spectrograph in a dedicated programme aimed at searching for radial velocity
substellar companions.
These stars compose the comparison sample (stars without known planets, SWOPs).
The total number of stars analysed in this work amounts to 70, from which
27 belongs to the SWP sample and 43 constitute the SWOPs. 
The stars are listed in table~\ref{basic_param_table}, whilst figure~\ref{hrdiagram}
shows the corresponding HR diagram.

\subsection{Observations}
The high-resolution spectra used in this work comes from two major sources, a
dedicated Nordic Optical Telescope DDT proposal (program 67-261, PI: E. Villaver) 
with the FIES spectrograph \citep{2014AN....335...41T} and from the ESO 
Archive\footnote{\url{https://archive.eso.org/eso/eso_archive_main.html}}. 
On the specific, HARPS \citep{2003Msngr.114...20M}, FEROS \citep{1999Msngr..95....8K} and ESPRESSO
\citep{2013Msngr.153....6P} spectra were used.
Additional HARPS-N \citep{2012SPIE.8446E..1VC} spectra of six stars were taken from
the TNG\footnote{\url{http://archives.ia2.inaf.it/tng/}} archive. 
Table~\ref{tab_spec} summarises the properties of the
different spectra. 

\begin{table}[!tb]
\centering
\caption{Properties of the different spectrographs used in this work.}
\label{tab_spec}
\begin{tabular}{lcrr}
\hline\noalign{\smallskip}
Spectrograph & Spectral range (nm) & Resolution & $N$ stars\\
\hline 
FIES      & 364-736 & 67000  &  9  \\
HARPS     & 378-691 & 115000 & 50  \\
HARPS-N   & 383-693 & 115000 &  6  \\
FEROS     & 350-920 & 48000  &  3  \\
ESPRESSO  & 380-788 & 140000 &  2  \\
\hline
\end{tabular}
\end{table}

All the spectra were reduced by the corresponding pipelines which 
implement the typical corrections involved in \'echelle spectra reduction.
When needed several spectra of the same star were properly combined.
The spectra were corrected from radial velocity shifts by using the precise radial
velocities provided by the pipelines.
Otherwise, radial velocities were measured by cross-correlating their spectra with spectra
of radial velocity standard stars of similar spectral types. 
In order to normalise the spectral regions of interest, we first used a percentile filter (set to 85\%)  to
identify the upper envelope of the flux, ignoring the absorption lines. Then, the continuum
was fitted iteratively to a second order polynomial using a sigma-clipping procedure.


\subsection{Spectroscopic analysis}\label{spec_analysis}
The effective temperature (T$_{\rm eff}$), surface gravity ($\log g$) and projected rotational velocity (v$\sin i$)
of each program star were determined by performing a simultaneous fit of the Balmer lines H$\beta$, H$\gamma$, and H$\delta$.
These lines are well known to be sensitive to the stellar parameters in AF-type stars \citep[e.g.][]{2005oasp.book.....G}.

A region of 10 \AA \space around the centre of each line, excluding the central 1.5 \AA  \space region was fitted.  
Solar metallicity, ATLAS9 atmospheric models \citep{2003IAUS..210P.A20C} were used for the calculations. 
In order to find the best suit of stellar parameters, we made use of a Bayesian fitting procedure.
The parameter space was sampled with {\tt emcee} \citep{Foreman-Mackey_2013}, 
based on the affine-invariant ensemble sampler for Markov chain Monte Carlo 
\citep[MCMC,][]{goodman2010ensemble}.
In addition to the stellar parameters we also allow for a small vertical
shift to account for uncertainties in the continuum normalisation. Uniform priors were used for 
T$_{\rm eff}$, v$\sin i$ and the vertical shift. A Gaussian prior were used for the surface gravity.
In order to get an initial estimate of $\log g$, we followed an iterative process. 
We start by running our MCMC code in order to get a first value of the effective temperature of the star.
Then, we computed
bolometric corrections using Gaia \citep{gaia_paper} colours\footnote{\url{https://gitlab.oca.eu/ordenovic/gaiadr3_bcg}}
\citep{2023A&A...674A..26C}, assuming a main-sequence star with solar composition.
We derived the stellar radius from the stellar luminosity and the effective temperature. Using a mass-luminosity
relationship, the mass of the star was computed. Surface gravity is then derived from mass and radius.
The process is repeated until the value of surface gravity converges (usually in two iterations). 

Our choice of a Gaussian prior for surface gravity is not arbitrary. It is well-known that for hot, fast-rotating
stars, an independent surface-gravity measurement is needed to uniquely identify the temperature from the Balmer lines
\citep[see e.g.][]{2005oasp.book.....G}. Otherwise, if $\log g$ is completely free, the MCMC might explore a parameter space
that is not physically possible, leading in some cases to $\log g$ values close to the prior upper limit, that in our
case was set to 5.0 (dex). This would have a severe impact on the derived temperatures and abundances.

\begin{figure}[htb]
\centering
\includegraphics[scale=0.60]{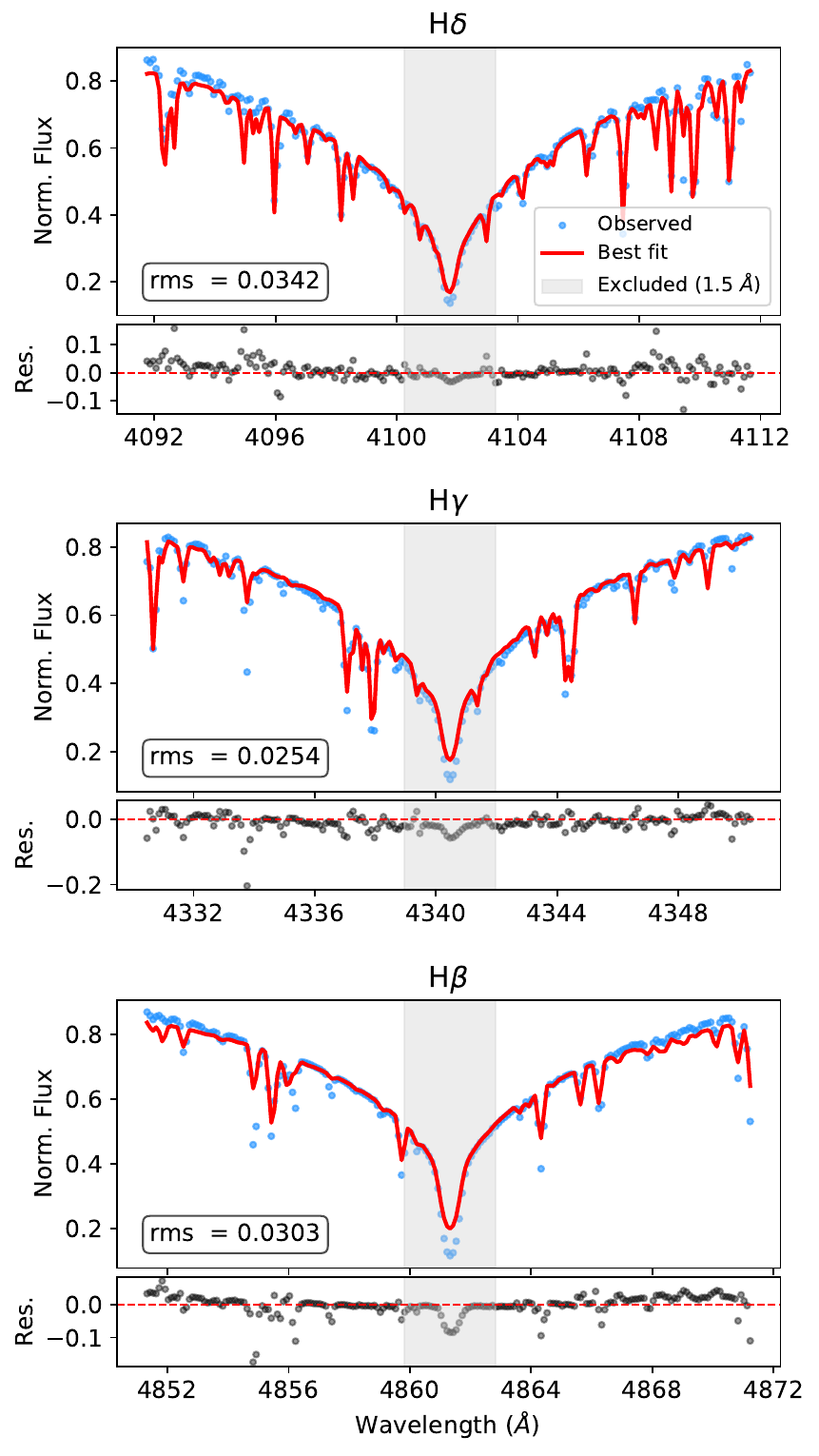}
\caption{Fit of the Balmer lines for the star TOI-1431. The blue points 
show the observed spectra while the best fit is provided as a red line.
The central 1.5 \AA \space core of the lines (marked as a grey shadow) is not taken into
account. For each line, H$\beta$, H$\gamma$, and H$\delta$, the residuals
are also shown. The rms (root-mean-squared) errors of the residuals are provided
in each subplot.}
\label{best_fit_example}
\end{figure}

The parameter space was covered with 32 walkers and we run the {\tt emcee} chain for 5000 steps.
The first values were eliminated by a burn-in phase, that was set as the first 1500 steps.
Each fit was carefully checked by computing the acceptance fraction and the autocorrelation time as
well as by a visual inspection of the corresponding fits, corner plots and trace evolution plots.
When needed, the {\tt emcee} analysis was repeated with a larger number of steps. 
An example of the derived fits is shown in figure~\ref{best_fit_example}, while figure~\ref{best_fit_corner} shows its corresponding corner plot. 
The full dataset of stellar parameters is provided in table~\ref{basic_param_table}.

Chemical abundances of individual elements were obtained by 
a multi-parameter fitting procedure based on the iterative Newton-Raphson method 
\citep[see e.g.][]{1992drea.book.....B}.
In order to apply this method to stellar spectra we followed the detailed description
of \cite{1995PASJ...47..287T}.
In brief, the method linearizes the dependence of the synthetic flux on the physical parameters
by computing a matrix of partial derivatives, allowing the code to optimise the parameters through
the inversion of the curvature matrix.
Uncertainties were formally estimated from the diagonal elements of the covariance matrix at the point of convergence.

In addition to the chemical abundances, we considered as
free parameters the rotational velocity, and additional Gaussian broadening to account for broadening
mechanisms different from rotation as well as a possible small radial velocity shift (to account for uncertainties on the
radial velocity correction). An important parameter in abundance determinations is the microturbulence velocity.
We computed it by using the analytical formula provided by \cite{2008JKAS...41...83T}.
Solar abundances were used as an initial guess. We also assumed an initial zero radial velocity shift,
while initial values of $v \sin i$ values were taken from the analysis of the Balmer's lines. 
Computations were done using ATLAS9 atmospheric models \citep{2003IAUS..210P.A20C}, whilst the atomic
data relevant to the analysis is taken from the extensive compilation of \cite{1995KurCD..23.....K} revisited by
\cite{2008JKAS...41...83T}. Synthetic spectra were computed at each step using the November 2019 version of the spectral synthesis code 
MOOG\footnote{\url{https://www.as.utexas.edu/~chris/moog.html}} \citep{1973PhDT.......180S} .

Departures from local thermodynamic equilibrium (LTE) can be pronounced in stars
with high temperature and with low gravity and metallicity. An exhaustive search for nLTE corrections in the literature
was performed and the corresponding corrections were applied as much as possible.
In particular, oxygen abundances derived from the O~{\sc i} 6158 $\AA$ lines were corrected from nLTE effects
using the nLTE corrections provided by \cite{2013AstL...39..126S};
the grid of nLTE corrections for calcium lines from \citet{2018MNRAS.477.3343S} was used to correct the corresponding abundances;
to correct the abundances derived from the Mg~{\sc i} b triplet lines, 
we made use of the calculations provided by \citet{2018ApJ...866..153A}.
A detailed list of available nLTE corrections for hot stars can be found in \citet{2020MNRAS.499.3706M},
including other elements such as C, and Ba. However, nLTE abundance analysis for these
elements is restricted to few stars which, in addition, are hotter (T$_{\rm eff}$ $>$ 9000 K) than the stars in our sample.

\begin{figure}[htb]
\centering
\includegraphics[scale=0.58]{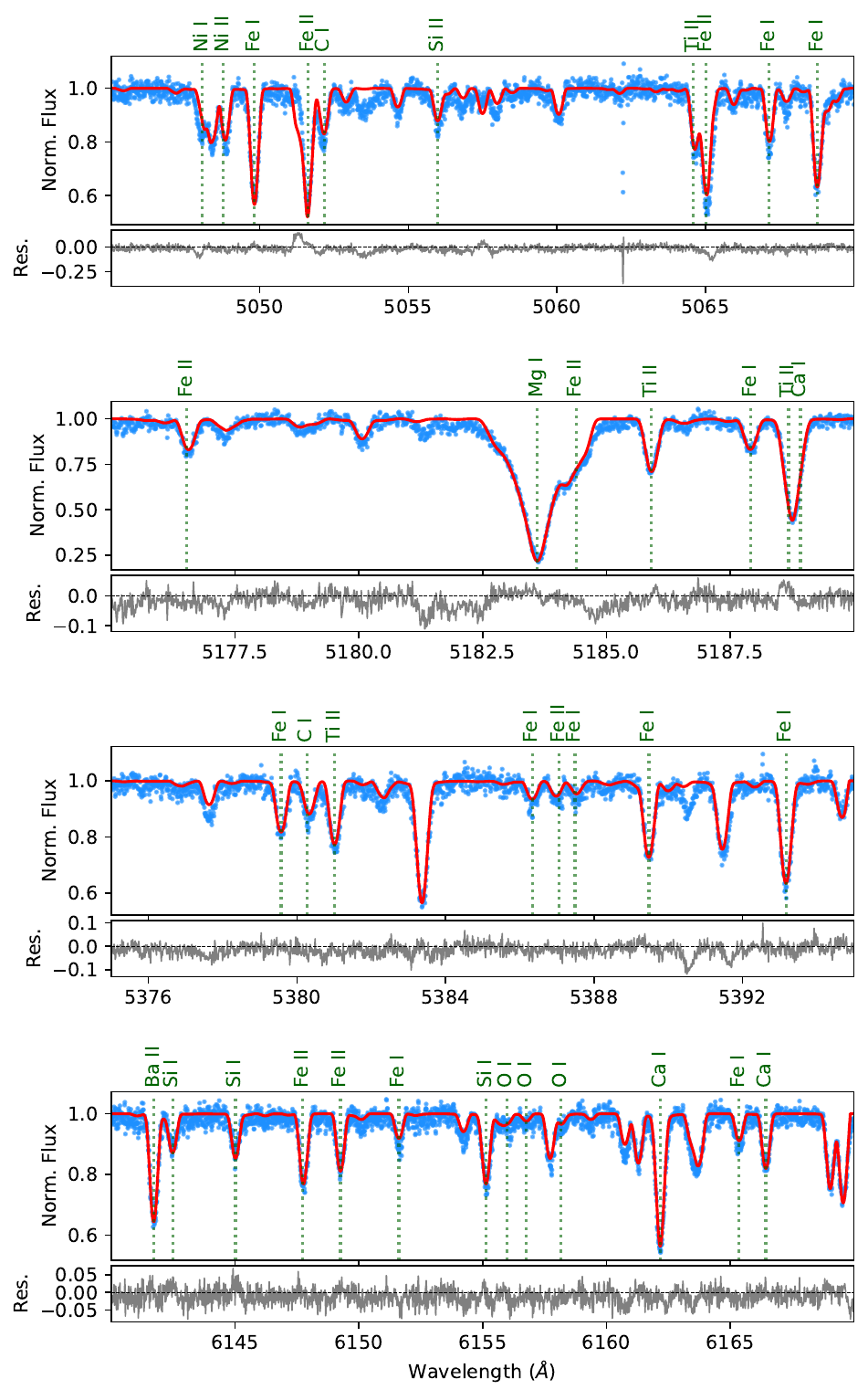}
\caption{Spectral fitting results for the star K2-334. The blue points
show the observed spectra whilst the best fit is shown as a red line.
For each fitted region, the residuals are also shown. The position of
several spectral lines is marked in dotted green lines.}
\label{best_fit_abundance}
\end{figure}

In the present study, we concentrated on four wavelength regions to be analysed.
These are listed in table~\ref{regions}. For each region we fitted for the abundances of the elements
that show lines of appreciable contribution.
Figure~\ref{best_fit_abundance} shows the theoretical synthetic spectra corresponding to the final solutions fit the
observations.
The derived abundances are listed in table~\ref{abd_param_table}.

Uncertainties in the derived abundances take into account three different contributions. The first is the formal uncertainty due to the spectral synthesis method. However, this statistical estimate does not account for other systematic errors such as atomic data, uncertainties in the atmosphere models, or continuum placement. A better estimate of the observational uncertainties was obtained for those elements measured in more than one spectral region by calculating the abundance dispersion between the different regions. 
In these cases, we compared this empirical dispersion with the propagated statistical uncertainty and adopted the larger of the two. For elements measured in a single spectral region, we took a conservative approach by doubling the method's uncertainty and adding a floor abundance error of 0.05 dex.
Furthermore, we also considered how the uncertainties on the stellar parameters affect the abundance determinations. To do this, three representative stars were selected, and abundances were re-derived while independently perturbing the standard values of the atmospheric parameters by $\pm 50$ K, $\pm 0.02$ dex, and $\pm 2$ km s$^{-1}$. The root-sum-square of the resulting abundance changes was then computed and incorporated as an additional source of uncertainty.
Final uncertainties were obtained by adding these independent sources of error in quadrature:
$\Delta_{\rm total}^{2} = \Delta_{\rm obs}^{2} + \Delta_{\rm stellar\_parameters}^{2}$, 
where $\Delta_{\rm obs}$ represents the observational error, defined as the maximum value between the synthesis fitting procedure uncertainty and the regional dispersion ($\Delta_{\rm obs} = \max(\Delta_{\rm fitting}, \Delta_{\rm dispersion})$).

\begin{table}[!tb]
\centering
\caption{Spectral regions analysed.}
\label{regions}
\begin{tabular}{cl}
\hline\noalign{\smallskip}
Spectral range (nm) & Elements\\
\hline
504.5 - 507.0 & C,  Si, Ti, Fe, Ni \\
517.5 - 519.0 & Mg, Ca, Ti, Fe \\
537.5 - 539.5 & C,  Ti, Fe \\
614.0 - 617.0 & O,  Si, Ca, Fe, Ba \\
\hline
\end{tabular}
\end{table}

\subsection{Discussion of potential biases}
The initial metal content of a star reflect the enrichment history of its birth
environment in the interstellar medium \citep[e.g.][]{1995ApJS...98..617T}. However, the photospheric abundances of 
early-F and A-type stars can be subsequently modified. In these fully or partially
radiative stars, surface abundances can be significantly altered by processes such
as planet engulfment or dust-gas segregation. 
Furthermore, to ensure that any observed chemical differences between our SWP and SWOP samples
are genuinely linked to planet formation, we must rule out biases related to Galactic
chemical evolution. Parameters such as age, kinematics and distance act as indirect tracers
of the Galactic populations where the stars were born \citep[e.g.][]{2013A&A...554A..84M,2020A&A...644A..68M}. 
For instance, metal-poor stars tend to belong to specific galactic populations (like the thick disc)
so they show particular spatial distributions.
Anticipating some results from the next sections, we notice that in the SWOP
sample there is a lack of stars with high abundance values.

Furthermore, our SWP and SWOP samples are subject to different observational selection biases. The SWOP sample relies heavily on Radial Velocity surveys, which typically select against hotter and evolved stars to avoid stellar jitter, focusing instead on intrinsically less luminous dwarfs. Conversely, the SWP sample includes mainly transiting planets from missions like {\sc Kepler} and {\sc TESS}. These transit surveys are less restricted by stellar jitter and are more heavily affected by Malmquist bias (especially in the {\sc Kepler} field), naturally incorporating a higher fraction of intrinsically luminous stars.

\begin{figure}[htb]
\centering
\begin{minipage}{0.49\linewidth}
\includegraphics[scale=0.425]{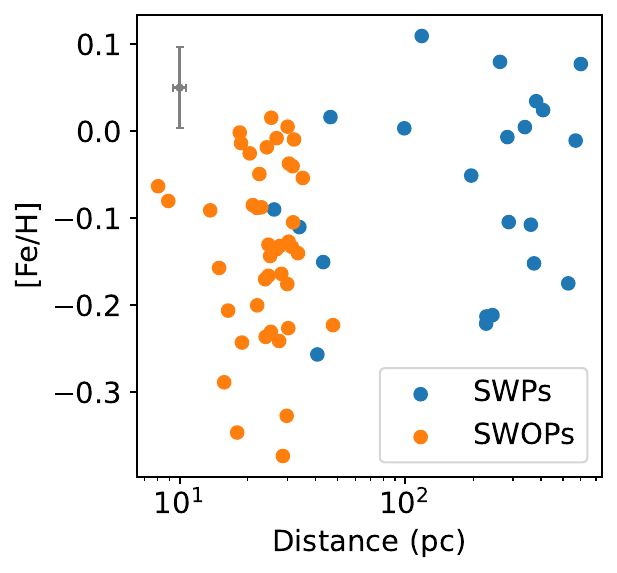}
\end{minipage}
\begin{minipage}{0.49\linewidth}
\includegraphics[scale=0.425]{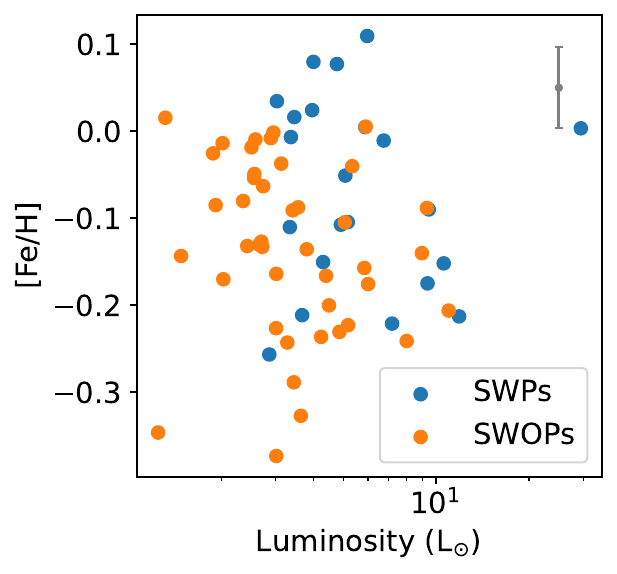}
\end{minipage}
\caption{
[Fe/H] as a function of the distance (left) and the luminosity (right) for the
stars in the SWOP and in the SWP samples.}
\label{metal_distance}
\end{figure}

Figure~\ref{metal_distance} (left) shows
the derived metallicity as a function of the stellar distance.
Distances are from the {\it Gaia} parallaxes \citep{gaia_paper}. 
In order to assess whether the metallicity of the star is distance-dependent, we first
compute the Spearman correlation test for SWPs and SWOPs. The results are compatible 
with a no correlation between metallicity content and distance.  For SWPs, we get
$r$ = 0.25, with a $p$-value of 0.26, whilst for SWOPs we 
obtain $r$ = 0.03, $p$-value = 0.86. In other words, the correlation distance-metallicity
is consistent with zero. 
Furthermore, we apply a multiple linear regression model to check whether or not the distance
can be use to predict the metallicity. In brief, the data is fitted with a model with the form
$ \rm{ [Fe/H]} = \beta_0 + \beta_1 \cdot \text{planetary state} + \beta_2 \cdot \log(\text{d})$,
where planetary state is one or zero (each star either has a detected planet or not).
We find that for the parameter
$\beta_2$, the $p$-value associated is  0.50, meaning that the distance has no predictive power on
the metal content. Table~\ref{ols_distance} gives the full details of the analysis.
The model metrics $R^{\rm 2}$ (proportion of variance explained) and $F$-statistics
(which measures the significance of the predictors)
are provided.

Even though SWOPs are brighter than SWPs, it is clear from Fig.~\ref{hrdiagram} that
SWPs are more luminous than SWOPs. Figure~\ref{metal_distance} (right) shows the metallicity
as a function of the luminosity. 
The figure reveals that there is no clear correlation between the two variables. 
A Spearman test confirms that for SWPs there is no correlation ($r$ = -0.16, with a $p$-value of 0.47),
whilst for SWOPs there might be a rather weak correlation ($r$ = -0.33, with a $p$-value of 0.03).
In principle, this is expected. At a given mass, the opacity in the interior of a more metallic star 
should be higher, making it more difficult for the photons to escape towards the stellar surface. 
A multiple linear regression model, see Table~\ref{ols_luminosity}, confirms  that the 
luminosity does not have a significant effect on the stellar metallicity.
The independent variables can only
explain the $\sim$ 10\% of the total metallicity variation. 
Whilst the parameter  $\beta_1$ is statistically significant (thus, recovering the planet-metallicity correlation)
the parameter $\beta_2$ is not ($p$-value =  0.23). 
Consequently, after controlling for the presence of planets, luminosity has no significant effect on stellar metallicity.
 
The physical mechanism relating opacity and luminosity should hold for both SWOPs and SWPs. The fact that we observe a possible weak correlation for SWOPs but none for SWPs can be an artefact of small-number statistics or related to subtle differences in mass, evolutionary stage, a restricted metallicity range or other criteria related to the selection of the stars in the SWP sample. 
In any case, our multiple linear regression analysis demonstrates that the differences in luminosity between SWPs and SWOPs do not significantly affect the overall comparison of their chemical abundances.

\begin{table}
\centering
\caption{Results of the ordinary least squares (OLS) regression model assessing the effect of planet presence and stellar distance on the metallicity.}
\label{ols_distance}
\begin{tabular}{lccc}
\hline
Predictor &  Value & Standard error & $p$-value \\
\hline
Intercept., $\beta_{\rm 0}$       & -0.097      & 0.034      & 0.006 \\
planet (yes), $\beta_{\rm 1}$     &  0.080      & 0.030      & 0.009 \\
$\log$(distance), $\beta_{\rm 2}$ & -0.069      & 0.057      & 0.231 \\
\hline
\multicolumn{4}{l}{Model metrics}\\
\hline
\multicolumn{4}{l}{$R^{\rm 2}$: 0.106}\\
\multicolumn{4}{l}{$F$-statistics ($p$-value): 3.669 (0.031)}\\
\hline
\end{tabular}
\end{table}

\begin{table}
\centering
\caption{Results of the ordinary least squares (OLS) regression model assessing the effect of planet presence and stellar luminosity on the metallicity.}
\label{ols_luminosity}
\begin{tabular}{lccc}
\hline
Predictor &  Value & Standard error & $p$-value \\
\hline
Intercept.,          $\beta_{\rm 0}$           &       -0.091 &     0.034  & 0.099 \\
planet (yes),        $\beta_{\rm 1}$           &        0.078 &     0.030  & 0.011 \\
$\log$($L_{\star}$), $\beta_{\rm 2}$           &       -0.081 &     0.057  & 0.159 \\
\hline
\multicolumn{4}{l}{Model metrics}\\
\hline
\multicolumn{4}{l}{$R^{\rm 2}$: 0.099}\\  
\multicolumn{4}{l}{$F$-statistics ($p$-value): 3.455 (0.038)}\\
\hline
\end{tabular}
\end{table}

Galactic-spatial velocity components $(U,V,W)$ were computed following the procedure
described in \cite{2010A&A...521A..12M} and the stars were classified as kinematic
members of the galactic thin / thick disc population following the method developed 
by \cite{2003A&A...410..527B,2005A&A...433..185B}. We find that all of our target stars 
show a kinematics compatible with being members of the thin disc population (with the only exception of
HD 91889 that shows an intermediate thin-thick disc kinematic).

The kinematic evolution of the Galaxy can also be used to derive stellar ages. 
Whilst this method might not be suitable to derive
precise ages for individual stars, it could be useful to compare different populations 
\citep{2010ARA&A..48..581S}.
To do this, we follow the Bayesian procedure depicted by \cite{2018MNRAS.476..184A} 
and \cite{2018ApJ...863..166V}.
For completeness, the corresponding age distributions as well as some
statistical diagnostics are provided in figure~\ref{ecdf_age} (left) and Table~\ref{ks_table_age}.
Both SWP and SWOPs show  an overall similar age distribution. Planet hosts have a mean age of  3.82 Gyr
with a standard deviation of 2.30 Gyr. On the other hand, the comparison sample
has a mean age of  4.62 Gyr with a dispersion of  2.71 Gyr. 

Finally, we have also compared the SWP and SWOP samples in terms of effective temperature,
finding no differences between both samples, see figure~\ref{ecdf_age} (right).
We conclude that our analysis demonstrates that there are no biases regarding distance, 
kinematics or age that might affect a comparison study of the abundances of
our samples of stars with and without known planets. 
It is clear, however, that we cannot rule out the presence of unknown planets in the SWOP sample.
In any case, the detailed detection limits computed by \cite{2009A&A...495..335L}
allow us to discard the presence of Jupiter-like planets in short orbits around these stars.
We are aware that, ideally, the SWOP should come from stars targeted for planetary transits
(as most SWPs do) but, unfortunately,
such stars are usually distant and too faint for high-resolution abundance studies.

It is well-known that some early-type stars show specific
abundance patterns, such as metallic-lined A-type or $\lambda$ Bo\"otis
stars. 
Peculiar Am stars are characterized by overabundances of most heavy elements in their spectra,
particularly Fe and Ni, together with under-abundances of Ca and Sc \citep[][and references therein]{2019MNRAS.484.2530C}.
On the other hand, $\lambda$ Bo\"otis stars show near-solar values of C and O, with sub-solar values of
the other metals \citep[e.g.][]{2002A&A...381..959H}.
We compared the abundance of our sample stars with the average pattern of
Am and $\lambda$ Bo\"otis stars and found two Am candidates
(namely KELT-17 and TOI-1431) as well as three possible 
$\lambda$ Bo\"otis stars (MASCARA 4, TOI-2109 and EPIC 246851721).
All these stars are planet hosts. 
As discussed in Sect.~\ref{introduction},
the relationship between chemically peculiar stars and planet hosts has
been studied in detail by \cite{2021A&A...647A..49S,2022A&A...668A.157S}. 
In the following, we will exclude these stars from the comparison of the chemical
abundances of SWPs and SWOPs. 
The corresponding abundance pattern of these stars
are shown in Figure~\ref{peculiar_stars}.

\section{Analysis}\label{analysis}
\begin{figure}[htb]
\centering
\includegraphics[scale=0.45]{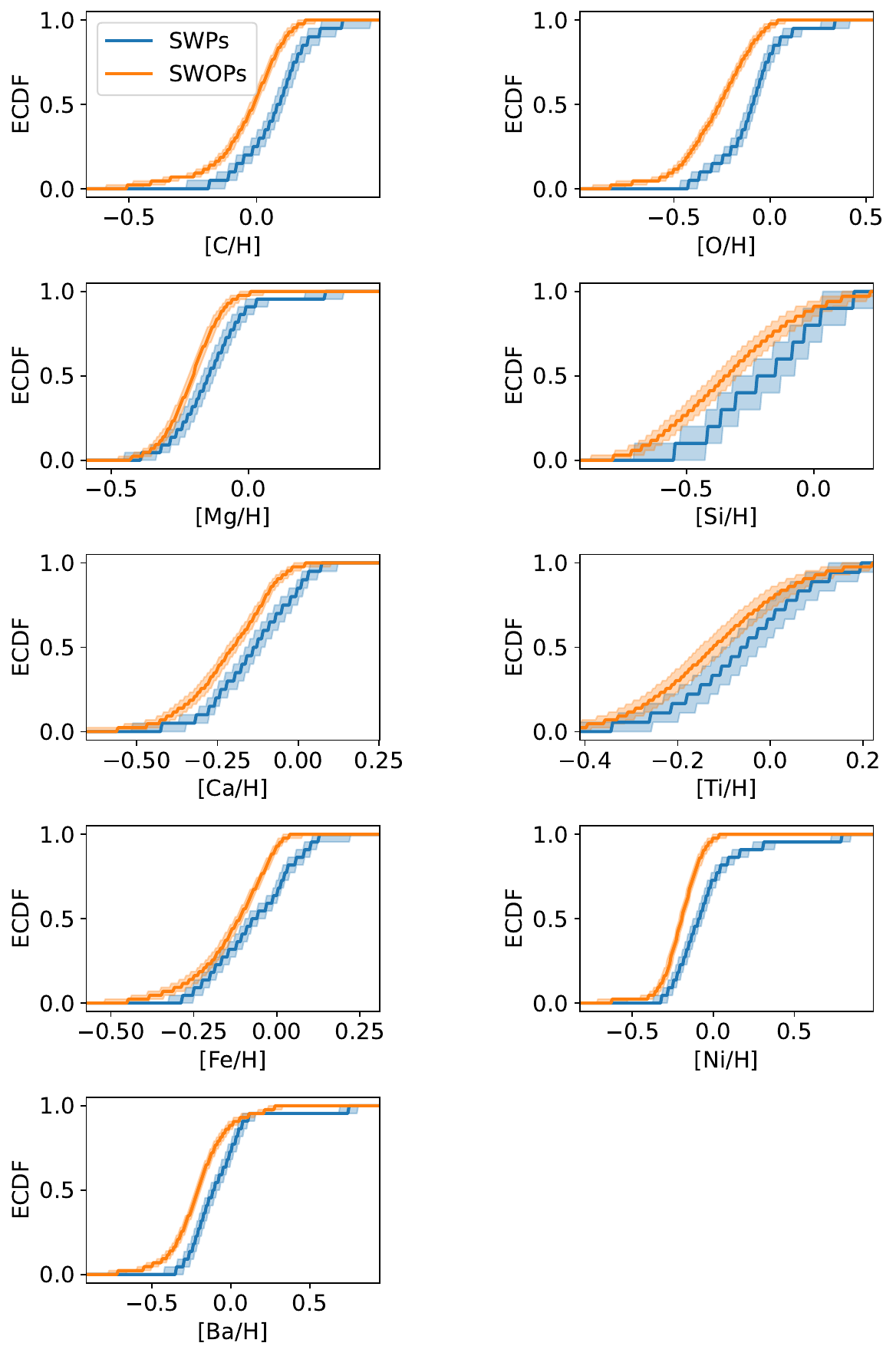}
\caption{Empirical cumulative distribution function (ECDF) of
[X/H] for SWPs (blue) and SWOPs (orange).}
\label{empirical_ecdf}
\end{figure}

Figure~\ref{empirical_ecdf} compares the empirical cumulative distribution function of the chemical abundances
of the SWP sample with those of the SWOPs. For guidance, some statistical diagnostics are also
provided in table~\ref{ks_table}.
There is a clear visual trend in most elements: SWPs tend to have abundance distributions shifted toward
higher values with respect to SWOPs. 
In order to assess whether or not the abundance distributions of SWPs and SWOPs are equal from a statistical point of view,
two statistical tests were performed: the Anderson-Darling \citep[A-D; e.g.][]{adtest}
and a standard two-sample Kolmogorov-Smirnov \citep[K-S; e.g.][]{1983MNRAS.202..615P}. 
We took into account the uncertainties in the derived abundances by performing a series of 10$^{\rm 4}$
simulations. In each simulation, abundances were randomly varied within their corresponding uncertainties before
computing the statistical tests.
Given that the probability distribution of the test statistics and their corresponding $p$-values might be not symmetric around its maximum,
we followed the common practice of reporting  
the median (50\% percentile) and
the range in parameters that delimits 68.2\% of the integrated distribution (percentiles 16\% and 84\%),
which is equivalent to the 1$\sigma$ limits for a Gaussian distribution.

\begin{table*}
{\scriptsize
\centering
\caption{Comparison between the abundances of SWPs and SWOPs.
}
\label{ks_table}
\begin{tabular}{lcccccccccc}
\hline
Element  & \multicolumn{3}{c}{SWOPs}   &  \multicolumn{3}{c}{SWPs}      & \multicolumn{2}{c}{K-S test} & \multicolumn{2}{c}{A-D test}      \\
         &  $N$  & Mean  &   $\sigma$  & $N$   &  Mean   &  $\sigma$    &  $D$          & $p$-value    &  $D$          & $p$-value         \\
\hline
C  & 43 & -0.04 & 0.14  & 20 &  0.08 & 0.10  & 0.417 (0.351, 0.491) & 0.011 (0.002, 0.051) & 4.930 (3.077, 7.026) & 0.004 (0.001, 0.018) \\
O  & 43 & -0.28 & 0.18  & 20 & -0.10 & 0.16  & 0.491 (0.434, 0.548) & 0.002 (0.000, 0.008) & 7.310 (5.619, 9.208) & 0.001 (0.001, 0.002) \\
Mg & 43 & -0.21 & 0.07  & 22 & -0.14 & 0.12  & 0.335 (0.260, 0.408) & 0.055 (0.009, 0.236) & 2.392 (0.761, 4.510) & 0.034 (0.005, 0.160) \\
Si & 34 & -0.33 & 0.15  & 10 & -0.20 & 0.14  & 0.371 (0.265, 0.476) & 0.187 (0.040, 0.566) & 0.838 (-0.326, 2.602) & 0.148 (0.028, 0.250) \\
Ca & 43 & -0.22 & 0.11  & 20 & -0.14 & 0.11  & 0.307 (0.251, 0.371) & 0.120 (0.034, 0.296) & 2.082 (0.791, 3.679) & 0.045 (0.011, 0.155) \\
Ti & 43 & -0.12 & 0.04  & 18 & -0.06 & 0.05  & 0.289 (0.198, 0.397) & 0.193 (0.026, 0.623) & 0.666 (-0.494, 2.914) & 0.175 (0.021, 0.250) \\
Fe & 43 & -0.13 & 0.10  & 22 & -0.07 & 0.11  & 0.317 (0.271, 0.375) & 0.075 (0.026, 0.179) & 2.435 (1.379, 3.624) & 0.033 (0.011, 0.088) \\
Ni & 43 & -0.20 & 0.10  & 22 & -0.04 & 0.23  & 0.423 (0.357, 0.477) & 0.008 (0.001, 0.034) & 5.822 (3.932, 7.934) & 0.002 (0.001, 0.009) \\
Ba & 43 & -0.20 & 0.18  & 22 & -0.07 & 0.21  & 0.353 (0.293, 0.420) & 0.042 (0.009, 0.120) & 3.009 (1.750, 4.513) & 0.019 (0.005, 0.062) \\
\hline
$\rm {C/O}$   &  43 & 1.05 & 0.18  & 18 & 0.89 & 0.26  & 0.249 (0.204, 0.314) & 0.253 (0.081, 0.484) & 1.735 (0.963, 2.740) & 0.063 (0.025, 0.131) \\
$\rm {Mg/Si}$ &  34 & 1.54 & 0.65  & 10 & 1.29 & 0.42  & 0.336 (0.336, 0.357) & 0.052 (0.034, 0.052) & 3.554 (3.098, 4.328) & 0.012 (0.006, 0.018) \\
\hline
\end{tabular}
\tablefoot{
For each subsample we list the number of stars ($N$) as well as the mean and the standard deviation values of
the [X/H] distribution. For the statistical tests, we provide
the test statistic value, $D$, as well as the asymptotic $p$-value.
Numbers in parenthesis correspond to the 16\% and 84\% percentiles of the distribution. 
}}
\end{table*}

The K-S analysis confirms that the abundance distributions of SWPs and SWOPs
are different for  C, O, Ni, and Ba ($p$-value $<$ 0.05). 
Furthermore, the differences between both subsamples are rather 
important: the maximum distance between the distributions ($D$ parameter) ranges
from $\sim$ 0.3 (Ba) to $\sim$ 0.5 (O).
Thus implies a systematic tendency of SWPs to show higher abundances than SWOPs. 
The A-D tests allow us to reject the null hypothesis of SWPs and SWOPs
coming from a similar parent distribution with a confidence level higher than 99.9\% in
most cases. This not only confirms the robustness of the systematic differences,
but also suggests that the divergence between the two populations remains 
strong even in the extremes (the tails) of the abundance distributions.

\begin{figure}[!htb]
\centering
\begin{minipage}{0.49\linewidth}
\includegraphics[scale=0.45]{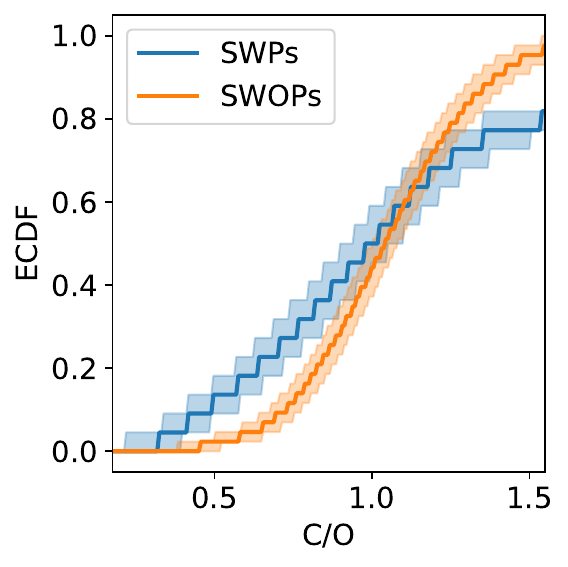}
\end{minipage}
\begin{minipage}{0.49\linewidth}
\includegraphics[scale=0.45]{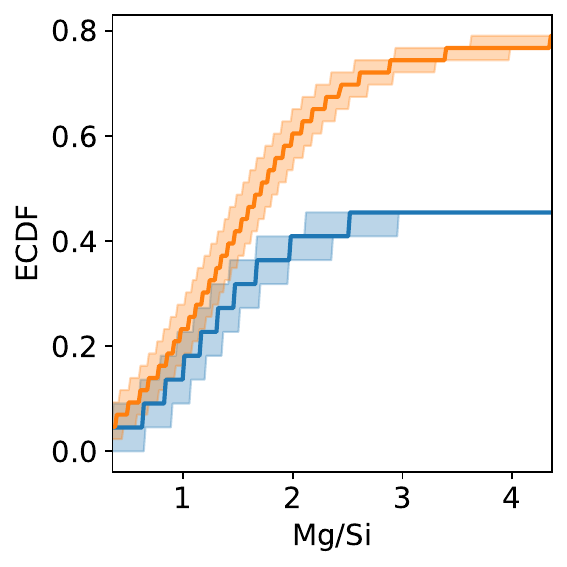}
\end{minipage}
\caption{
 $\rm {C/O}$ (left) and  $\rm {Mg/Si}$ (right) ratio cumulative fraction of SWPs (blue) and SWOPs (orange).}
\label{empirical_ecdf_ratios}
\end{figure}

In order to set our results into a wider context, we now discuss how the frequency of
planets depend on the abundance of different groups of elements. More specifically,
we classify the elements into volatiles (C, O), $\alpha$-elements (Mg, Si, Ca, Ti)
and iron-peak  (Ni, Fe).

For the analysis, we followed the Bayesian procedure described
in detail in \citep[e.g.][]{2010PASP..122..905J,2025A&A...695A..27M}. 
In those works, the fraction of stars with planets was fitted with
a function of the form $f[X/H] = C \cdot 10^{\beta \cdot [X/H]}$
by following a Bayesian approach.
However, as previously discussed,
at high abundance values we only have stars in the SWP sample.
This might affect our analysis by an effect known as the quasi-complete separation problem \citep{10.1093/biomet/71.1.1}. As we do not have
SWOPs at high values, the only possibility that the MCMC algorithm has to
maximise the likelihood is to artificially increase the strength of the planet-abundance
relationship. 
To overcome this problem, we replace the fitting function form with a logistic function
with the functional form

\begin{equation}
\label{metaleq}
f([X/H]) = \frac{C \cdot 10^{\beta [X/H]}}{1 + C \cdot 10^{\beta [X/H]}}
\end{equation}

\noindent where [X/H] is the mean abundance of the different groups of elements.
At low-abundances the denominator is $\sim$ 1, so the equation behaves as the widely assumed $f[X/H] = C \cdot 10^{\beta \cdot [X/H]}$.
On the other hand, at high-abundances the function goes smoothly to 1, avoiding unrealistic probabilities and fitting the observations.
We impose Gaussian priors on the $\beta$ parameter, $\mathcal{N}$ ($\mu$ =2, $\sigma$ = 1.5) for iron-peak elements, and  $\mathcal{N}$ ($\mu$ =0, $\sigma$ = 5) in the case of the $\alpha$ and
volatile elements. For the $C$ constant, in all cases, a truncated-normal distribution was considered
$\mathcal{N}_{[0,+\infty]}$ ($\mu$ = 0.05, $\sigma$ = 0.1).

The derived mean $\beta$ values and their corresponding uncertainties are
$\beta$ = 1.565 $\pm$  0.771  for the iron-peak elements,
$\beta$ = 2.689 $\pm$  1.314  for volatile elements,
and  $\beta$ = 0.063 $\pm$ 0.792  when [X/H] is the mean $\alpha$-elements abundance.
For completeness, the marginal posterior probability is shown in Fig.~\ref{corner_plots}.

\section{Discussion}\label{discussion}
\subsection{A differential chemical fingerprint: Iron-peak, volatiles and $\alpha$-elements}

As shown, the abundance distributions of SWPs  seem to be shifted towards higher abundance values than SWOPs, although this tendency is only statistically significant for a few elements.
These results fit well in the framework of core-accretion models for planet formation.
The core accretion model for giant planet formation begins with planetesimal coagulation forming a massive core of roughly 5 - 20 M$_{\oplus}$ \citep[e.g.][]{2018exha.book.....P}.
In this scenario, a high-metallicity environment implies a high dust-to-gas-ratio in the protoplanetary disc that facilitates condensation, and accelerates accretion before the gas
disc is lost \citep[e.g.][]{1996Icar..124...62P,2003ApJ...598L..55R,2004A&A...417L..25A,2012A&A...541A..97M}.
It was proposed early on that Jupiter-size planets form at relatively low temperatures beyond the snowline
and that hot Jupiters experience subsequent inward migration to their current locations
\citep{1996Natur.380..606L}.
Metal-rich protoplanetary discs do not only facilitate the formation of a primordial core, but are also
more massive and dynamical thus, favouring disc migration
\citep{2008IAUS..249..331M,2010MNRAS.402.2735E}.
Within this context, carbon and oxygen are key elements to form ices \citep[e.g.][]{2011ApJ...743L..16O}, so
larger abundances of volatiles imply a significant amount of solid material at large distances
from the host stars.
However, it should be noted that the in situ formation through core accretion of hot Jupiters has also been
discussed \citep[e.g.][]{2016ApJ...829..114B,2018A&A...612A..93M,2024A&A...686L...1M}.
Finally, it is also important to take into account that we are focusing on A and early-F type stars
which show high levels of high-energy radiation that might erode their protoplanetary discs
in relatively short timescales. Indeed, protoplanetary discs dissipate faster as the stellar mass increases, although
the underlying processes are not fully understood \citep{Hernandez_2005,2015A&A...576A..52R}.
Therefore, only in high-metallicity environments can a rocky core form rapidly before gas dissipation occurs.

We also find that 
different groups of elements 
trace planet formation in distinct ways. While the absolute mass contribution of certain trace metals
is small compared to elements like C, O, Mg, Si or Fe, their stellar abundances 
might correlate with planetary occurrence. 
For instance,
there seems to be no dependence on the abundance of pure rocky material (i.e., $\alpha$-elements), whilst 
the planetary occurrence rate shows a dramatically steeper exponential dependence on iron-peak elements,
in agreement with the findings for FGK main-sequence stars \citep[e.g.][]{2005ApJ...622.1102F}.
We show that a similar dependency is found for volatile elements.
Thus, the limiting factor for the formation and survival of these giant planets is not simply
the production of silicate material, but rather the efficient accumulation of heavy metals and ices. 


\subsection{Migration and formation environment}

Stellar  $\rm {C/O}$ and  $\rm {Mg/Si}$ ratios are known to play an important role in constraining the planetary composition.
The  $\rm {C/O}$ ratio alters the abundances of carbon grains and water ices, thus 
shifting the location of key snowlines \citep[e.g.][]{2021ApJ...909...40T}.
On the other hand, the  $\rm {Mg/Si}$ value controls the silicate mineralogy and dictates whether Mg is mainly
in the form of pyroxene (MgSiO$_{\rm 3}$) or olivine (Mg$_{\rm 2}$SiO$_{\rm 4}$) \citep[e.g.][]{2010ApJ...715.1050B,2015A&A...577A..83D}.
We should caution that our current sample includes mostly gas-giant planets, for which
the role of the stellar $\rm {Mg/Si}$ ratio in their mineralogy is still far to be fully understood.
Along these lines, it is worth noticing that \citet{2015A&A...581L...2A} showed that after removing the
Galactic evolution trend, low-mass planet host stars show higher [Mg/Si] ratios, but
stars without planets and stars with gas-giant planets show similar [Mg/Si] ratios.

We compared the  $\rm {C/O}$ and  $\rm {Mg/Si}$  ratios for SWP and SWOPs in Fig.~\ref{empirical_ecdf_ratios}; see also table~\ref{ks_table}.
Our statistical analysis reveals that the $\rm {C/O}$ ratios for both populations are broadly similar.
Although there is a visual tendency for SWOPs to exhibit slightly
higher $\rm {C/O}$ values in the range $\rm {C/O}$ $<$ 1.0,
the K-S test yields a $p$-value of 0.25, whilst the A-D test gives $p$ = 0.06. This indicates no strong
statistically significant difference. This lack of a clear relative excess of carbon in SWPs
suggests that late-stage stellar pollution by carbon-enriched material from planet migration is not a dominant mechanism
altering the host star's initial $\rm {C/O}$ ratio.

On the other hand, the $\rm {Mg/Si}$ ratio
shows a statistically significant difference between the two samples 
(K-S $p$ = 0.05, A-D $p$ = 0.01), with SWPs tending towards higher $\rm {Mg/Si}$ ratios
compared to SWOPs at high ratio values. 
This might be explained by variations in the primordial cloud composition or by selective condensation
and sequestration of specific rocky materials during planet formation.

\subsection{Stellar evolution and chemical fingerprints of planet formation} 
\cite{2015A&A...582L..10K} showed that Herbig stars with transitional discs having
radial cavities or gaps show a deficit of refractory elements, and
thus low values of metallicity. This finding was confirmed by \cite{2023A&A...671A.140G} 
who found that Herbig stars with transitional discs tend to have lower
metallicities but strong (sub-)millimetre continuum emission likely
related to gas-giant planets. 
\cite{2015A&A...582L..10K} and \cite{2018MNRAS.476.4418J} developed a framework in which
forming planets trap the metal-rich material, while metal-depleted material
continues to flow towards the central star.
More recently, \cite{2025A&A...695A..27M} tracked the metallicity content from
the pre-MS to the red giant phase of intermediate-mass (M$_{\star}$ $>$ 1.5 M$_{\odot}$) stars.
They found that the
differences in metal content between planet and non-planet MS hosts were rather modest and the strength of the planet-metallicity correlation was significantly lower than for the less massive FGK MS stars. For stars on the red-giant branch, a strong planet-metallicity correlation, compatible with that found for FGK MS, stars was found.
As intermediate-mass stars are mainly radiative, the metallicity of the star does not reflect its bulk composition but the composition of the accreted material during planet formation. When the star leaves the MS and develops a sizeable convective envelope, a strong-planet metallicity correlation is recovered.

\begin{figure}[htb]
\centering
\includegraphics[scale=0.45]{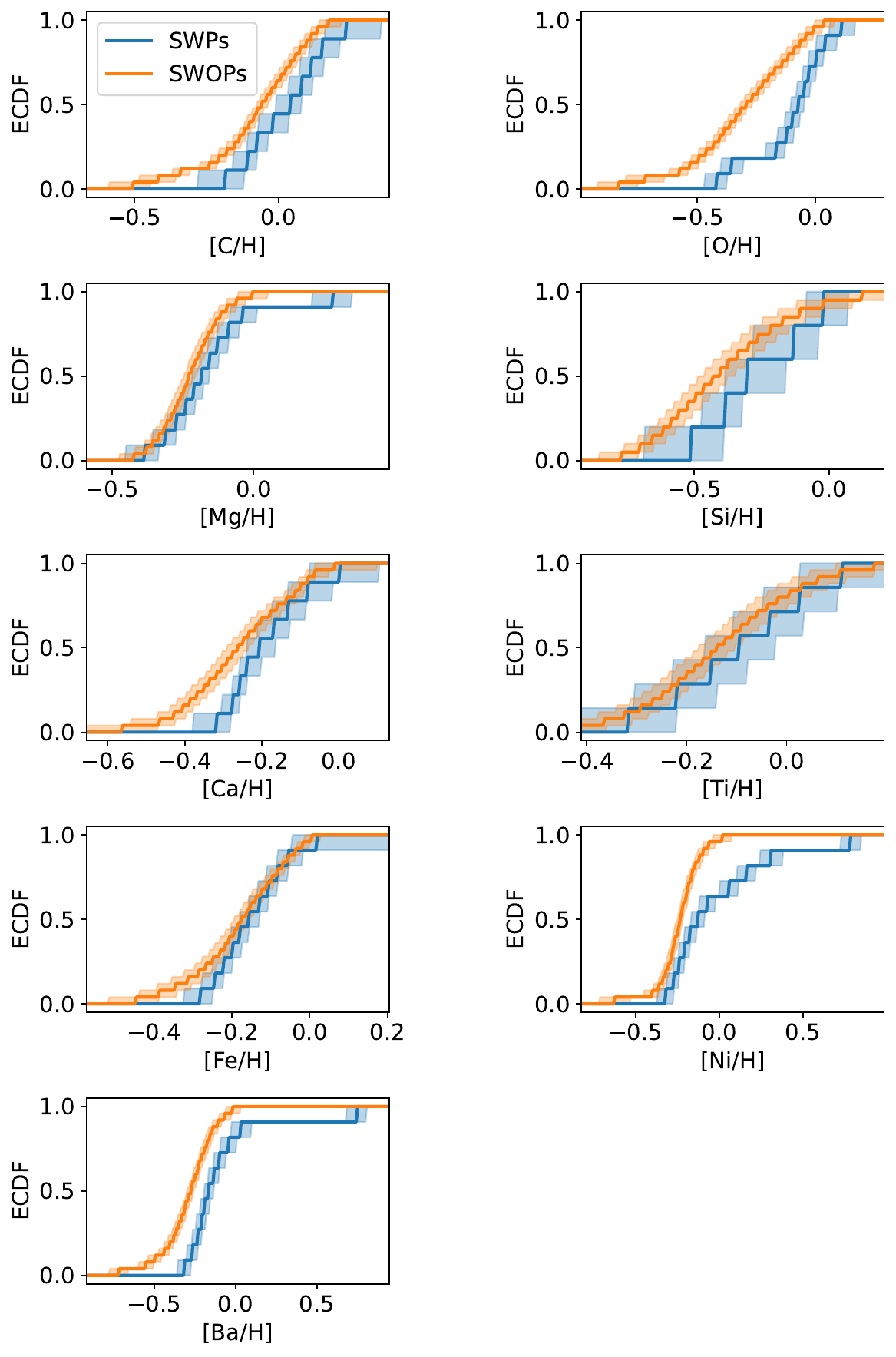}
\caption{ 
[X/H] cumulative fraction of SWPs (blue) and SWOPs (orange) when only stars with spectral type earlier than F5 are considered.}
\label{empirical_ecdf_full_radi}
\end{figure}

Our sample of planet hosts contains A and early-F type stars, so we have a mixture
of still convective and fully radiative stars. For a proper comparison with the aforementioned
works, it is therefore mandatory to select those stars with effective temperatures higher than 6650 K, that is, spectral types
earlier than F5, which we assume to be fully radiative. The results are shown in figure~\ref{empirical_ecdf_full_radi}, where the distribution of abundances for SWPs and SWOPs
are shown and in table~\ref{ks_table_radiative} that summarises the statistical tests.
It can be seen that the trend of SWPs to show larger abundances than SWOPs remains, however 
there are several differences with respect to the previous analysis that 
need to be discussed.

\begin{table*}
{\scriptsize
\centering
\caption{ 
Same as Table~\ref{ks_table}, but considering only stars with spectral type earlier than F5.}
\label{ks_table_radiative}
\begin{tabular}{lcccccccccc}
\hline
Element  & \multicolumn{3}{c}{SWOPs, SpType < F5}   &  \multicolumn{3}{c}{SWPs, Sp-Type < F5}      & \multicolumn{2}{c}{K-S test} & \multicolumn{2}{c}{A-D test}   \\
         &  $N$  & Mean  &   $\sigma$  & $N$   &  Mean   &  $\sigma$    &  $D$          & $p$-value    &  $D$          & $p$-value \\
	 \hline
C  & 25 & -0.08 & 0.16  & 9  &  0.03 & 0.11  & 0.369 (0.284, 0.467) & 0.251 (0.078, 0.559) & 0.777 (-0.214, 2.104) & 0.157 (0.044, 0.250) \\
O  & 25 & -0.31 & 0.21  & 11 & -0.11 & 0.14  & 0.538 (0.487, 0.618) & 0.015 (0.003, 0.035) & 3.903 (2.750, 5.216) & 0.009 (0.003, 0.024) \\
Mg & 25 & -0.23 & 0.08  & 11 & -0.16 & 0.16  & 0.298 (0.225, 0.396) & 0.410 (0.135, 0.752) & 0.064 (-0.598, 1.198) & 0.250 (0.104, 0.250) \\
Si & 20 & -0.39 & 0.16  & 5  & -0.27 & 0.12  & 0.450 (0.300, 0.600) & 0.337 (0.085, 0.826) & 0.066 (-0.714, 1.569) & 0.250 (0.073, 0.250) \\
Ca & 25 & -0.26 & 0.11  & 9  & -0.19 & 0.07  & 0.400 (0.298, 0.480) & 0.181 (0.064, 0.491) & 0.691 (-0.211, 1.896) & 0.171 (0.054, 0.250) \\
Ti & 25 & -0.14 & 0.04  & 7  & -0.10 & 0.05  & 0.337 (0.240, 0.474) & 0.457 (0.127, 0.839) & -0.179 (-0.784, 1.136) & 0.250 (0.111, 0.250) \\
Fe & 25 & -0.18 & 0.09  & 11 & -0.15 & 0.08  & 0.258 (0.200, 0.335) & 0.586 (0.287, 0.847) & -0.385 (-0.788, 0.259) & 0.250 (0.250, 0.250) \\
Ni & 25 & -0.24 & 0.11  & 11 & -0.01 & 0.31  & 0.455 (0.378, 0.527) & 0.058 (0.018, 0.171) & 3.075 (1.881, 4.660) & 0.018 (0.005, 0.054) \\
Ba & 25 & -0.29 & 0.15  & 11 & -0.08 & 0.27  & 0.509 (0.425, 0.600) & 0.025 (0.004, 0.092) & 3.712 (2.304, 5.408) & 0.010 (0.002, 0.037) \\
\hline
$\rm {C/O}$   & 25 & 1.02 & 0.20  & 9 & 0.77 & 0.22  & 0.367 (0.276, 0.465) & 0.195 (0.050, 0.510) & 1.229 (0.316, 2.422) & 0.101 (0.033, 0.250) \\  
$\rm {Mg/Si}$ & 20 & 1.72 & 0.78  & 5 & 1.49 & 0.49  & 0.345 (0.345, 0.396) & 0.255 (0.135, 0.255) & 1.575 (1.194, 2.237) & 0.073 (0.039, 0.105) \\
\hline
\end{tabular}
}
\end{table*}

\begin{figure}[!htb]
\centering
\begin{minipage}{0.49\linewidth}
\includegraphics[scale=0.45]{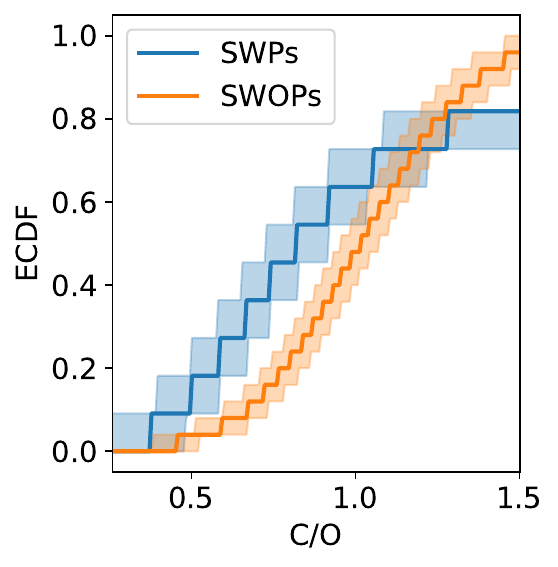}
\end{minipage}
\begin{minipage}{0.49\linewidth}
\includegraphics[scale=0.45]{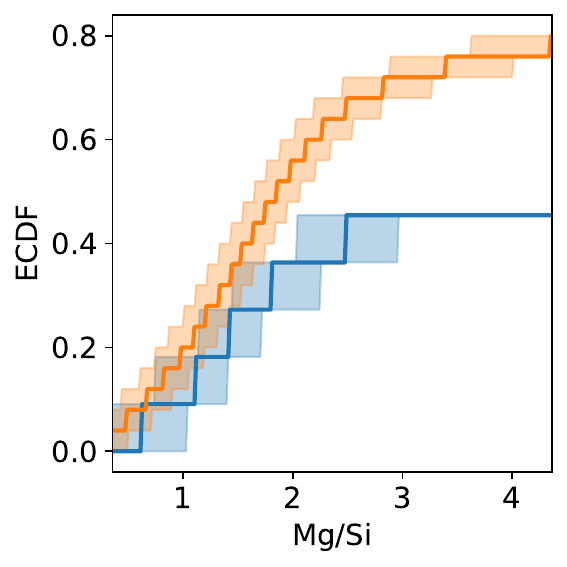}
\end{minipage}
\caption{
Same as Fig.~\ref{empirical_ecdf_ratios} but for fully radiative stars.}
\label{radiative_empirical_ecdf_ratios}
\end{figure}

{To start with, the behaviour of oxygen and carbon are decoupled. 
SWPs show larger oxygen abundances than SWOPs. 
This difference is well funded by the statistics with a K-S $p$-value = 0.015 and a A-D $p$-value = 0.009.
On the other hand, the abundance distributions of carbon for SWPs and SWOPs are
statistically similar. 

For refractory elements (Si, Fe, Mg, Ca, Ti) the mean abundances are higher in SWPs than in SWOPs and a visual
tendency of SWPs to have larger abundances is clearly visible. However, this tendency is not statistically significant, but
there are two clear 
exceptions, Ni (A-D $p$ = 0.018) and Ba (A-D $p$ = 0.010).
We should caution that statistical tests are very sensitive to sample size, therefore, it is more
difficult to prove that a visual difference is real and not an effect of statistical noise in a smaller sample.
The stringent sample selection criteria and rigorous error propagation limit the statistical power of the analysis for individual elements owing to the small sample size.
Regarding the $\rm {C/O}$ and the $\rm {Mg/Si}$ ratios (figure~\ref{radiative_empirical_ecdf_ratios})
there is a marginal tendency of SWPs to have lower $\rm {C/O}$. Conversely, although the arithmetic mean of the $\rm {Mg/Si}$ ratio appears mathematically lower in SWPs due to an extended tail in the SWOPs distribution (driving a low A-D $p$-value), their distribution curves reveal that the bulk $\rm {Mg/Si}$ composition remains statistically 
comparable (the K-S $p$-value remains high) between both populations.

In brief, SWPs show larger absolute abundances of O and some refractories like Ni and Ba, while their C abundance and bulk $\rm {Mg/Si}$ ratio are indistinguishable from those of SWOPs.
Silicates and oxides from asteroidal and planetesimal material are inherently rich in oxygen and refractory elements, 
yet they condense in a way that preserves the primordial, bulk-stellar $\rm {Mg/Si}$ proportion
\citep[e.g.][]{2003ApJ...591.1220L}.
The empirical evidence of such dry, rocky debris accretion (characterized by high oxygen
and refractory abundances but depleted carbon) is well documented in polluted white dwarfs
\citep[e.g.][]{2003ApJ...584L..91J,2016NewAR..71....9F}.
This disfavours the accretion of gas giants or distant comets (carbon-rich material), 
and suggests and origin tied to asteroids or fragments of terrestrial planets driven toward the star by dynamical instabilities within the system.

}


In order to further test the rocky accretion scenario, a Bayesian hierarchical analysis
of the abundance versus condensation temperature ([X/H]-$T_{\rm C}$) trend was performed.
Values of $T_{\rm C}$ correspond to a 50\% equilibrium condensation
temperature for a solar system composition gas \citep{2003ApJ...591.1220L}. 
When evaluating our full sample, the difference in the [X/H]-$T_{\rm C}$ slope between
SWPs and SWOPs is marginally  negative ($\Delta$-slope = -0.008 (dex / 1000 K)). 
When isolating the fully radiative stars, the mean difference in the [X/H]-$T_{\rm C}$ slope
between planet hosts and the comparison sample shifts towards a  positive 
$\Delta$-slope = 0.024 (dex / 1000 K), with the posterior probability
yielding a
69.5\% probability that SWPs exhibit a steeper positive dependence
on T$_{\rm C}$ than SWOPs, see Fig.~\ref{tc_abundance_posterior}.
While the rather small sample size of the radiative SWPs limits the formal statistical significance,
this pronounced steepening of the  [X/H]-$T_{\rm C}$ trend in the absence of convective mixing agrees
with the steady-state surface pollution by refractory-rich, volatile-poor debris.

Early F and A-type stars will eventually evolve into subgiants and red giant stars.
\cite{2013A&A...554A..84M}
showed that giant stars with planets (in the mass range M$_{\star}$ $>$ 1.5 M$_{\odot}$) show higher metallicities and nickel abundances than
a comparison sample of giant stars without known planets, while no differences were found between giant stars with and
without known planets in the abundance distribution of silicon and manganese.
This points towards a primordial origin of the planet-metallicity correlation, showing that early-F and A-type stars with planets
are born in a metal-rich environment.
On the other hand, if the higher abundances of O  and Ni that we see in these stars are due to the accretion of 
rocky material, this signature would be lost as the star evolves and the external layers are gradually diluted when the convective zone penetrates the envelope.
Red giants will end their stellar life as white dwarfs, many of which 
are enriched with traces of photospheric metals in their atmosphere
due to the disruption and subsequent accretion of asteroids \citep[e.g.][]{2003ApJ...584L..91J}.

Finally, we note that regarding the dependence of the planetary fraction on the abundance of
different types of elements, the results do not change when only radiative stars are considered.
The Bayesian fit of Eq.~\ref{metaleq} provide similar results to the previous case
($\beta$ = 1.197 $\pm$ 0.787  for the iron-peak elements,
$\beta$ =  0.735 $\pm$ 1.117  for volatile elements,
and  $\beta$ = -0.398 $\pm$ 0.938 when [X/H] is the mean $\alpha$-elements abundance,
see Fig.~\ref{corner_plots_f5}). 
We note that in \cite{2025A&A...695A..27M} only metallicity (instead of the mean abundance of iron-peak elements)
was considered, finding a value of $\beta$ = 0.80 $\pm$  0.60 for main-sequence, intermediate-mass stars.
In our sample of radiative early-F and A stars, if only metallicity is considered, a value of
$\beta$ = 0.023 $\pm$  0.936 is derived. Although both values are still compatible within errors, 
in our sample, $\beta$ is considerably lower. 
In the framework by \cite{2015A&A...582L..10K} this can be easily explained.
As intermediate-mass stars are mainly radiative, the metallicity of the star does not reflect its bulk composition but the composition of the accreted material
during the pre-MS phase.
The slow, non-convective
mixing motions in radiative stars are not able to fully recover the planet-metallicity correlation in this kind of stars.

\section{Conclusions}\label{conclusions}

In this work a detailed comparison of the chemical abundances of a sample
of early-F and A-type stars with and without known planets has been performed.
We show that for most of the considered elements, there is a visual tendency of  SWPs to show higher abundances than SWOPs,
supporting core-accretion models for planet formation and highlighting the key role of
elements other than iron on planet formation. 
However, for most elements this trend lacks statistical confidence.
The analysis of larger samples along with a better understanding of the atmospheres of early-type stars
including detalied nLTE effects
will help us to confirm or reject these apparent trends.

When the sample is restricted to fully radiative stars (spectral type earlier than F5),
carbon abundances and the mineralogical ratios $\rm {C/O}$ and $\rm {Mg/Si}$ of SWPs are not statistically different from that of SWOPs. This,
combined with a higher abundance of elements like O and Ni in SWPs, supports the 
accretion of silicate-rich, volatile-poor material that continuously pollutes the radiative
external layers of the host stars.

These results might provide a crucial missing link in the chemical life-cycle of planetary systems.
During the pre-main sequence phase, forming giant planets trap metal-rich dust in the protoplanetary disc,
temporarily leaving the host star depleted of refractory elements \citep[e.g.][]{2015A&A...582L..10K}.
Once the star reaches the main sequence our analysis suggests that dynamical instabilities driven by the mature planets throw rocky planetesimals
back into the star, creating a steady-state surface pollution of silicates.
Later, these stars evolve into the red giant phase, where the development of deep convective envelopes
dilutes this shallow superficial pollution, allowing the star to show its metal-rich primordial bulk composition \citep{2013A&A...554A..84M}.
The mechanism of dynamically-driven debris accretion which we observe starting in early-type main-sequence stars mirrors
the end-of-life processes seen in polluted white dwarfs, where surviving planetary architectures continue to disrupt and
accrete asteroid material \citep[e.g.][]{2002ApJ...572..556D,2003ApJ...584L..91J,2016NewAR..71....9F}.
Thus, the chemical fingerprints of early-type stars not only reveal the architecture of their planetary systems, but also act as a tracer of
their dynamical evolution from birth to death. 

Finally, we emphasise the importance of an accurate abundance characterisation of stellar hosts. To maximise detection probabilities, future searches for exoplanets around intermediate-mass stars should prioritise targets displaying the chemical pattern identified in this work: metal-rich stars with elevated abundances of both volatile and silicate-forming elements.

%
\bibliographystyle{aa}
\bibliography{early_hosts_chem.bib}


\begin{acknowledgements}
J. M. acknowledges support from the Italian Ministero dell'Università e della Ricerca and
the European Union - Next Generation EU through project PRIN 2022 PM4JLH ``Know your little neighbours: characterising low-mass stars and planets in the Solar neighbourhood''.
I.M. is funded by grant PID2022-138366NA-I00, by the Spanish Ministry of Science and Innovation/State Agency of Research MCIN/AEI/10.13039/501100011033 and by the European Union.
G.M.M. acknowledges financial support from Junta de Andaluc\'ia through the
program Emergia (EMEC\_2023\_00533).
B. M. acknowledges the funding by grant PID2021-127289-NB-I00 from MCIN AEI/10.13039/501100011033/ and FEDER. 
We sincerely appreciate the careful reading of the manuscript and the constructive comments of an anonymous referee.
This work is based on observations collected at the Nordic Optical Telescope through a DDT proposal
(program 67-261, PI: E. Villaver) as well as from observations from the public ESO Archive, program IDs
0100.A-9018(A), 0100.C-0097(A), 0100.C-0474(A)
0100.C-0750(A), 0100.C-0847(A), 0101.A-9008(A)
0102.A-9006(A), 0102.A-9008(A), 0102.A-9029(A)
0102.C-0584(A), 0103.C-0206(A), 0104.A-9012(A)
0104.C-0413(A), 0104.C-0418(A), 0108.A-9029(A),
072.C-0488(E),
072.D-0707(A),
073.C-0733(A,C,D,E),
073.C-0733(E),
074.C-0364(A),
075.C-0234(B),
075.C-0637(A),
075.C-0689(A,B),
076.C-0073(A),
076.C-0279(A,B,C),
076.D-0103(A),
077.C-0295(A,B,C,D),
078.C-0044(A),
078.C-0209(A,B),
078.C-0233(B),
079.C-0170(B),
080.C-0032(A),
080.C-0664(A),
080.C-0712(A),
081.C-0774(A),
082.C-0412(A),
083.C-0794(A,B,C,D),
084.C-1039(A),
085.C-0019(A),
087.C-0831(A),
088.C-0353,
089.C-0006,
089.C-0732(A),
089.C-0739(A),
090.C-0421(A),
094.A-9012(A),
094.C-0946(A),
094.D-0596(A),
095.C-0799,
095.C-0799(A),
096.C-0238(A),
096.D-0402(A),
097.C-0390(B),
098.C-0042(A),
098.C-0151(A),
098.C-0269(A,B),
098.C-0292(B),
098.C-0304(A),
098.C-0739(A),
099.C-0205(A),
099.C-0303(A),
099.C-0898(A),
105.20FX,
105.20N0.001,
106.20ZN.001,
106.20ZN.002,
106.20ZN.003,
106.21ER.001,
106.21TJ.001,
109.22Z4.006,
1101.C-0557(A),
1102.C-0923(A),
1102.C-0923(C),
182.D-0356(B),
183.C-0972(A),
184.C-0815(A,B,C,E,F),
192.C-0224,
192.C-0224(B,C,H),
198.C-0169(A),
60.A-9036(A),
60.A-9700(G),
60.A-9709(G).
Additional data from the Telescopio Nazionale Gaileo archive under programms
A26TAC\_70,
A26\_TAC70,
CAT19A\_97,
ITP19\_1 and GAPS were also used. 
\end{acknowledgements}

\begin{appendix}
\onecolumn
\section{Additional material}\label{appendix}

\begin{figure*}[!htb]
\centering
\includegraphics[scale=0.65]{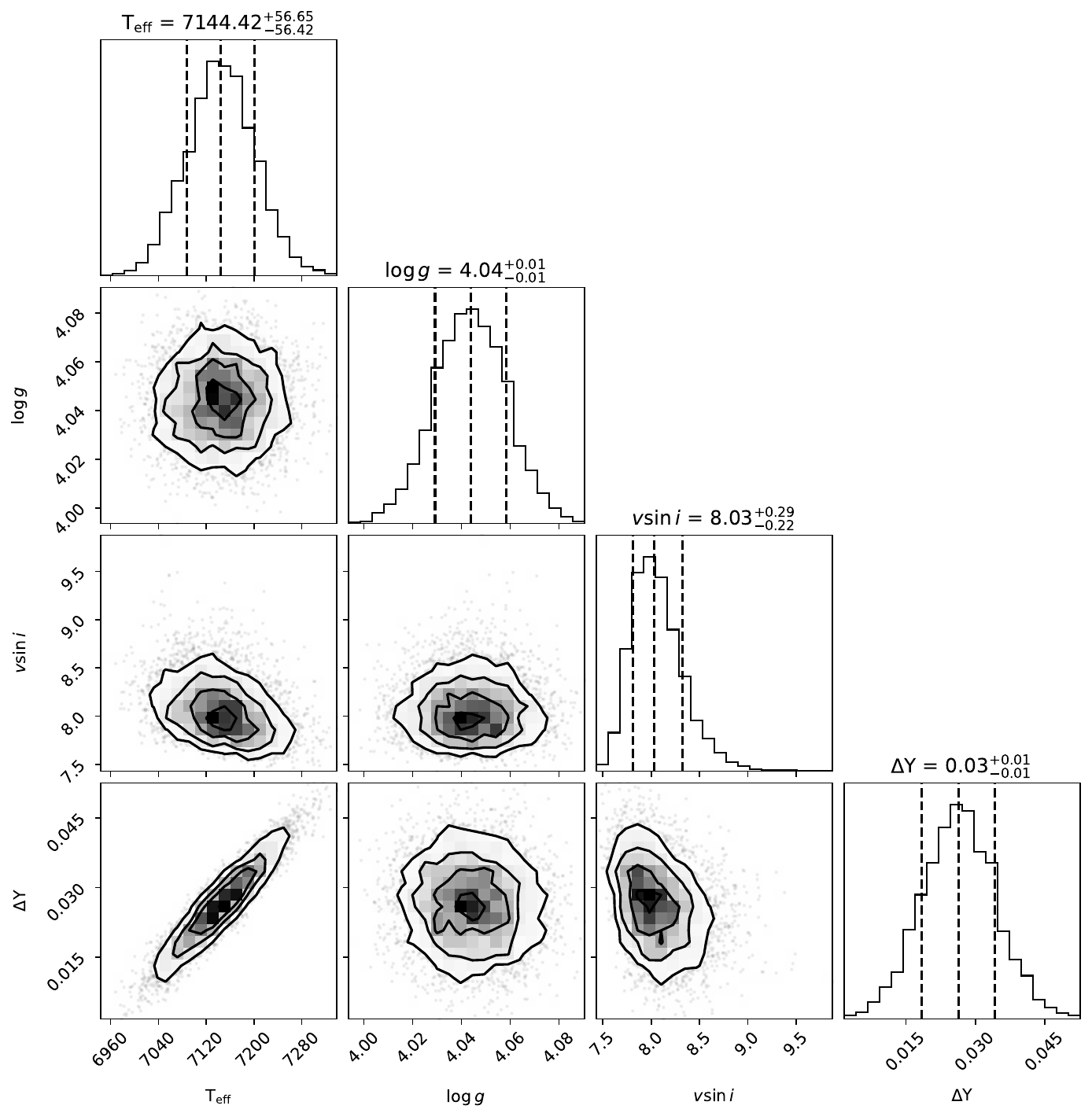}
\caption{Corner plot of the fit of the Balmer lines for the star TOI-1431.
}
\label{best_fit_corner}
\end{figure*}

\begin{table*}[!htb]
\centering
{\scriptsize
\caption{Comparison between the ages an effective temperature of SWPs and SWOPs.}
\label{ks_table_age}
\begin{tabular}{ccccccccccc}
\hline
Parameter  & \multicolumn{3}{c}{SWOPs}   &  \multicolumn{3}{c}{SWPs}      & \multicolumn{2}{c}{K-S test} & \multicolumn{2}{c}{A-D test}   \\
            $N$  & Mean (Gyr) &   $\sigma$ (Gyr) & $N$                 &  Mean (Gyr)   &  $\sigma$ (Gyr)   &  $D$          & $p$-value    &  $D$          & $p$-value \\
\hline
Age (Gyr)         & 43 & 4.62 & 2.71  & 22 & 3.82 & 2.30  & 0.212 (0.152, 0.286) & 0.463 (0.147, 0.820) & -0.203 (-0.754, 0.921) & 0.250 (0.137, 0.250) \\
T$_{\rm eff}$ (K) & 43 & 6749 & 237   & 22 & 6745 & 434   & 0.222 (0.180, 0.266) & 0.389 (0.200, 0.641) &  0.200 (-0.204, 0.698) & 0.250 (0.170, 0.250) \\
\hline
\end{tabular}
\tablefoot{
For each subsample we list the number of stars ($N$) as well as the mean and the standard deviation values of
the [X/H] distribution. For the statistical tests, we provide
the test statistic value, $D$, as well as the asymptotic $p$-value.
Numbers in parenthesis correspond to the 16\% and 84\% percentiles of the distribution.
}
}
\end{table*}

\begin{figure*}[]
\centering
\begin{minipage}{0.49\linewidth}
\includegraphics[scale=0.65]{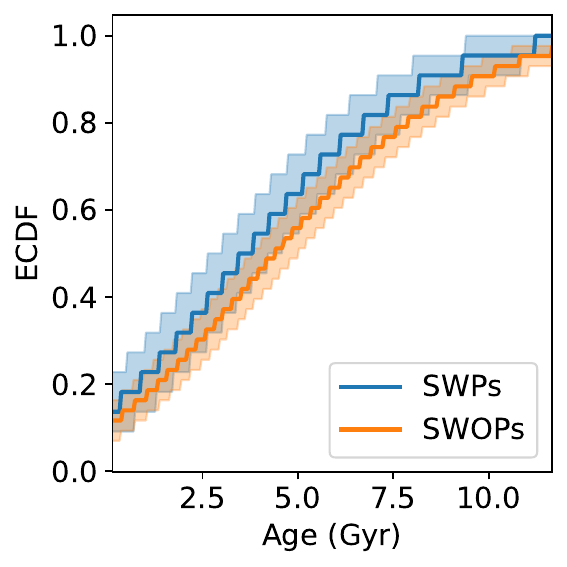}
\end{minipage}
\begin{minipage}{0.49\linewidth}
\includegraphics[scale=0.65]{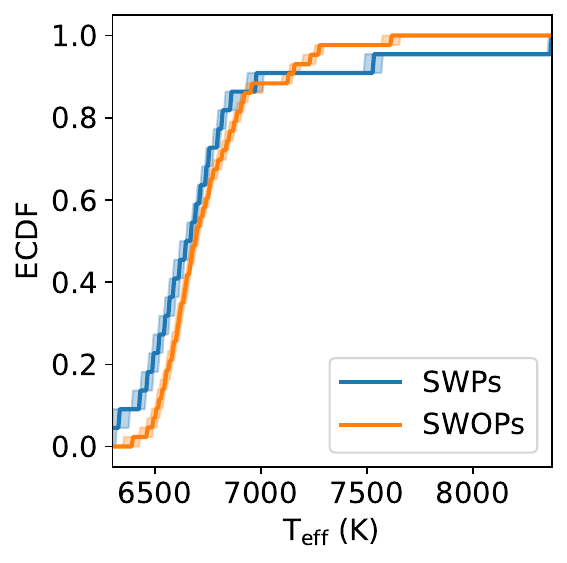}
\end{minipage}
\caption{ 
Kinematic age (left) and effective temperature (right) cumulative distribution of SWPs (blue) and SWOPs (orange).
}
\label{ecdf_age}
\end{figure*}

\begin{figure*}[htb]
\centering
\includegraphics[scale=0.60]{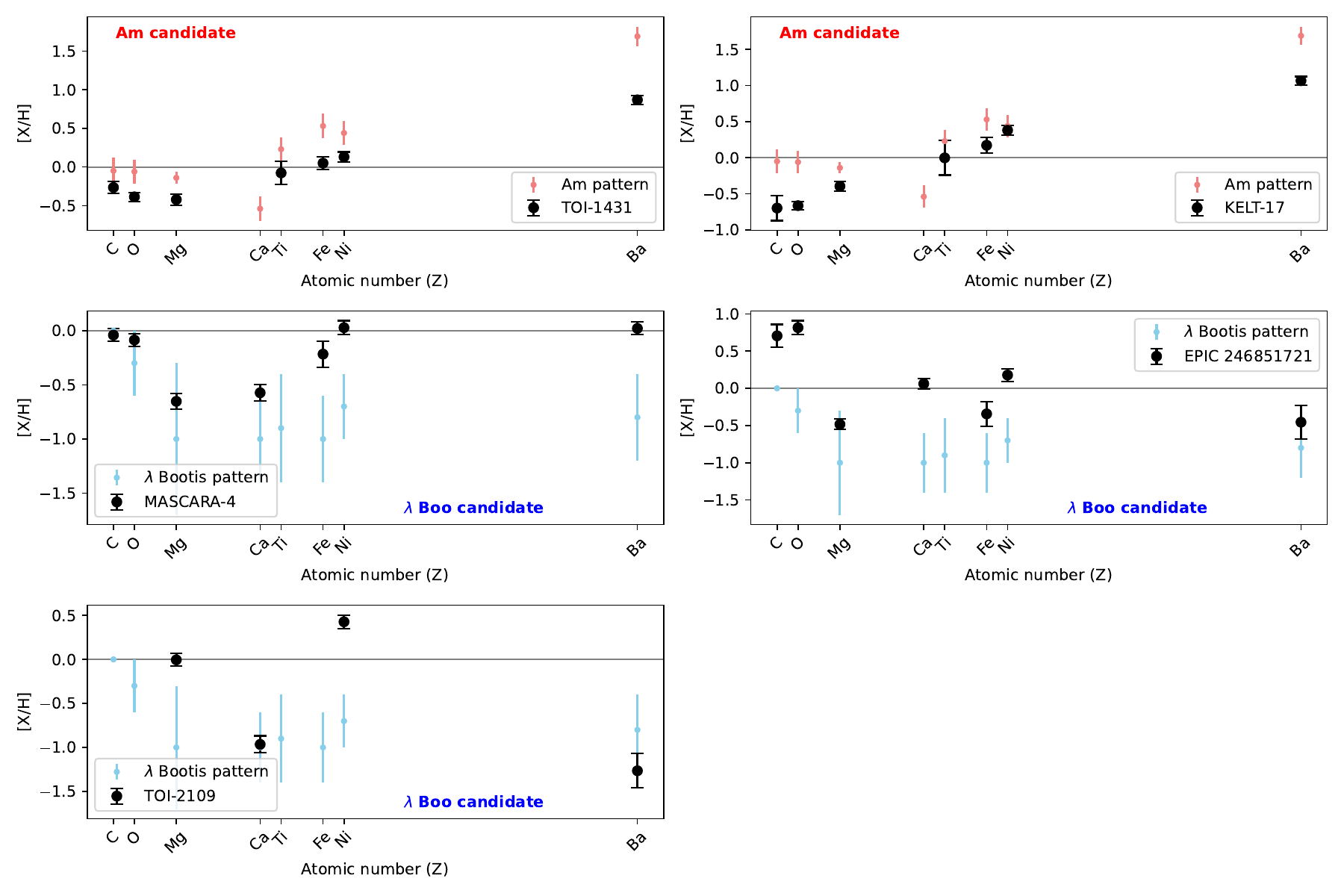}
\caption{ 
Abundances of TOI-1431 and KELT-17 compared to an average Am-star patter
\citep{2019MNRAS.484.2530C} and MASCARA 4, EPIC 246851721 and TOI-2109 compared to an average
$\lambda$ Bo\"otis pattern \citep{2002A&A...381..959H}.
}
\label{peculiar_stars}
\end{figure*}

\begin{figure*}[]
\centering
\begin{minipage}{0.33\linewidth}
\includegraphics[scale=0.40]{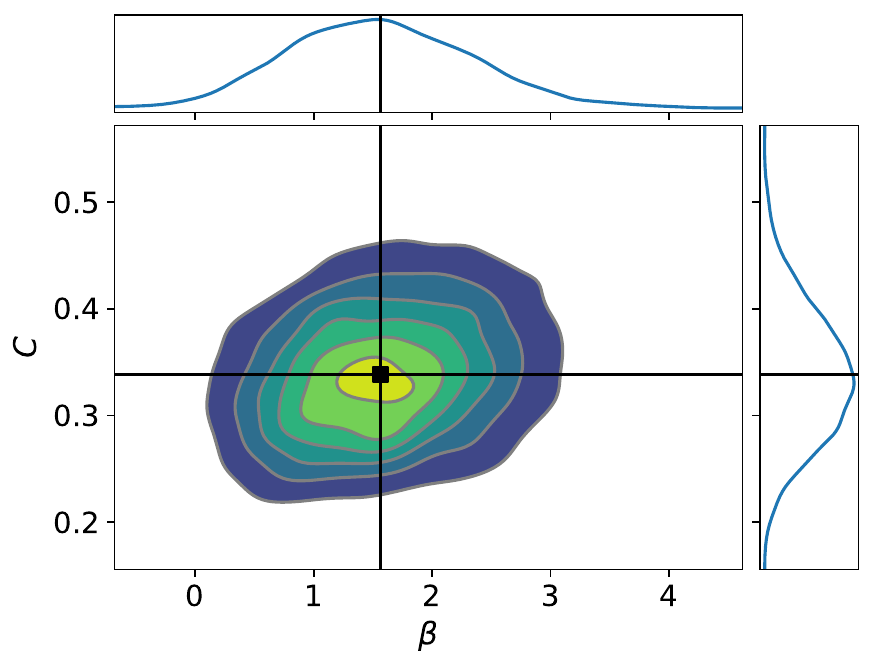}
\end{minipage}
\begin{minipage}{0.33\linewidth}
\includegraphics[scale=0.40]{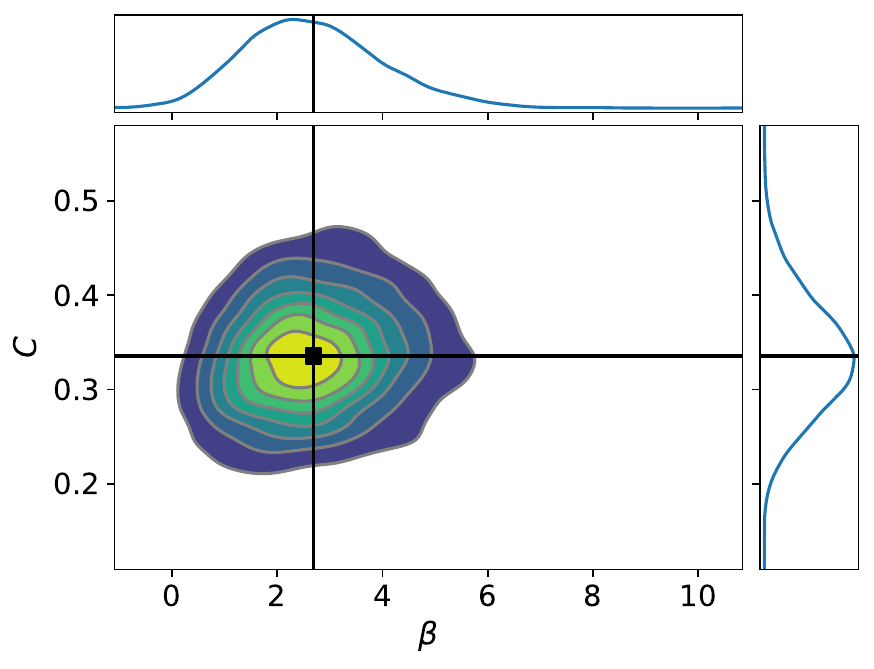}
\end{minipage}
\begin{minipage}{0.33\linewidth}
\includegraphics[scale=0.40]{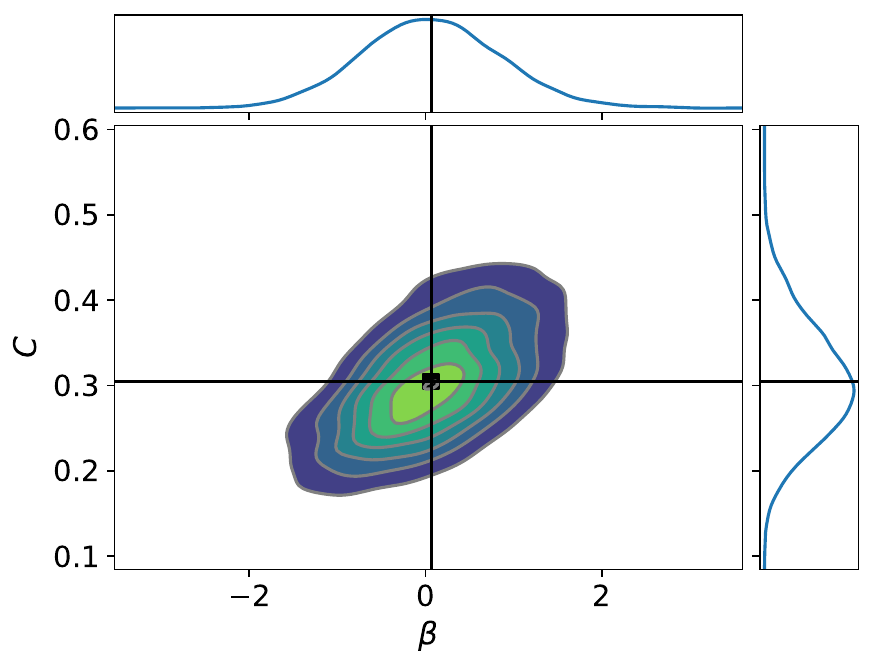} 
\end{minipage}
\caption{ 
Marginal posterior probability distribution functions of the Bayesian fit
to Eq.~\ref{metaleq} when [X/H] is the mean abundance of
iron-peak elements (left), volatile elements (middle), and $\alpha$ elements (right).
The vertical line indicates the mean of the distribution.
}
\label{corner_plots}
\end{figure*}

\begin{figure*}[]
\centering
\begin{minipage}{0.33\linewidth}
\includegraphics[scale=0.40]{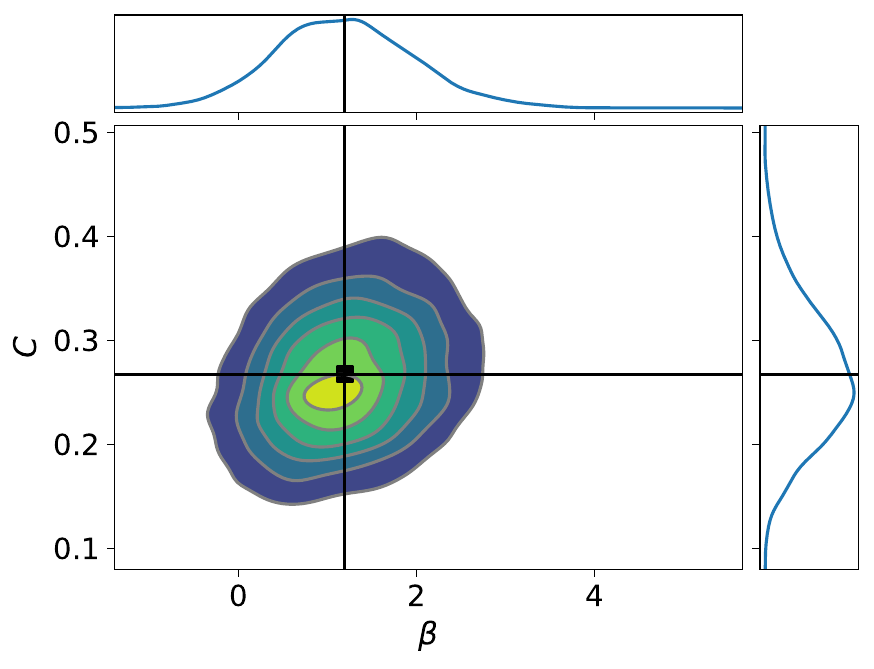} 
\end{minipage}
\begin{minipage}{0.33\linewidth}
\includegraphics[scale=0.40]{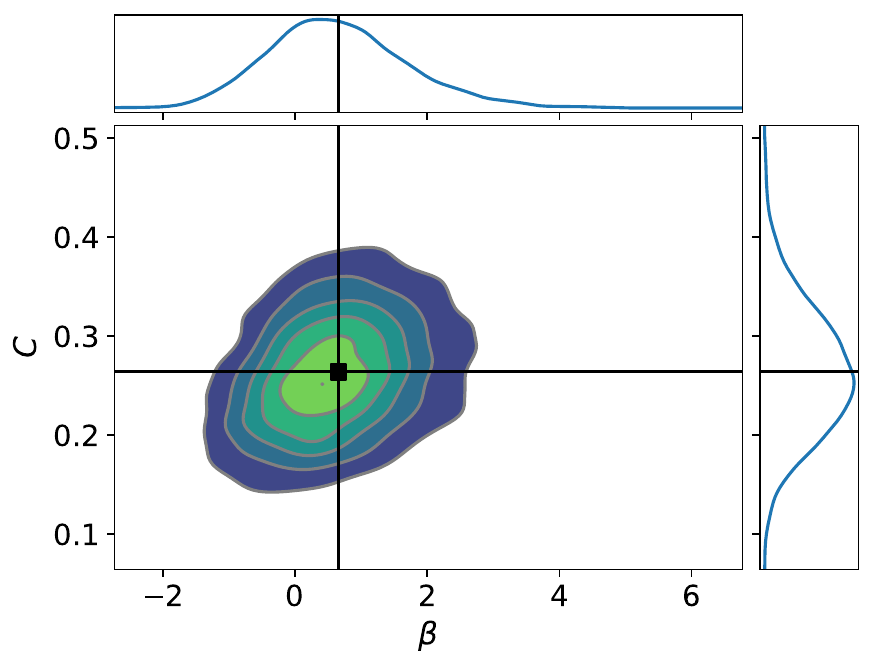}
\end{minipage}
\begin{minipage}{0.33\linewidth}
\includegraphics[scale=0.40]{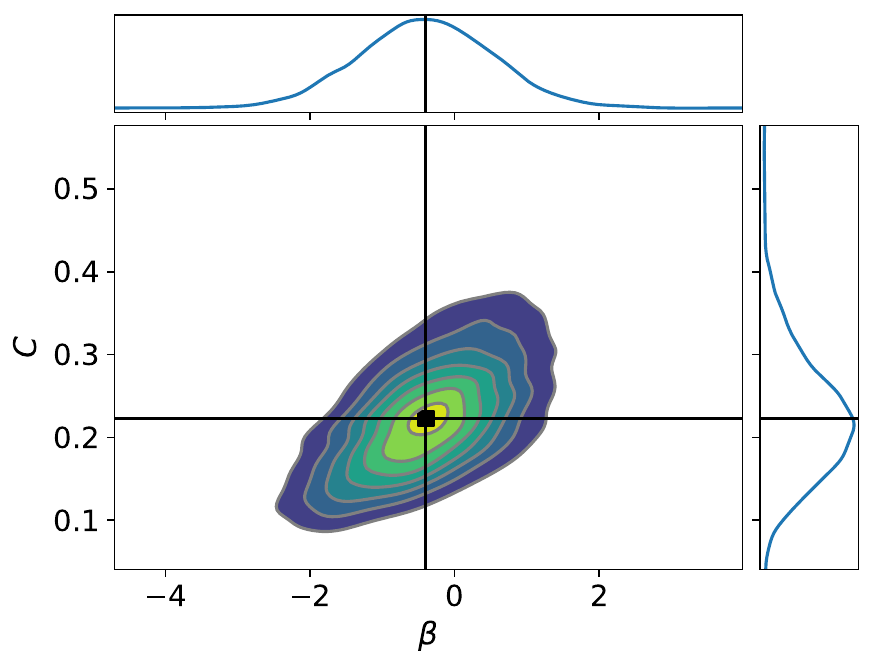} 
\end{minipage}
\caption{Same as Fig.~\ref{corner_plots} but considering only the
stars with a spectral type earlier than F5.}
\label{corner_plots_f5}
\end{figure*}

\begin{figure*}[htb]
\centering
\includegraphics[scale=0.45]{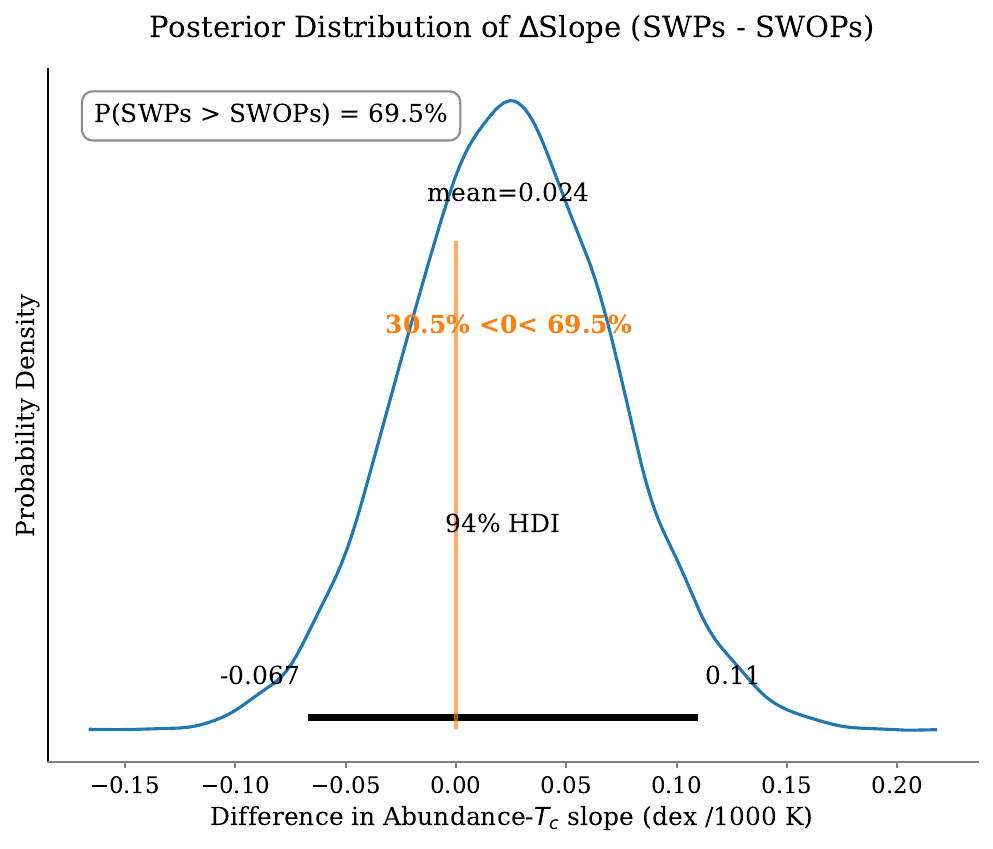}
\caption{ 
Marginal posterior probability distribution of the difference in the [X/H]-T$_{\rm C}$ slopes ($\Delta$-slope) between planet hosts (SWPs) and the comparison sample (SWOPs) for the fully radiative subsample (spectral types earlier than F5). The vertical line indicates the reference value of zero. The mean of the distribution and the 94\% Highest Density Interval (HDI) are
also shown.
}
\label{tc_abundance_posterior}
\end{figure*}

\onecolumn
\begin{longtable}{lcccccccccc}
\caption{\label{basic_param_table} Basic stellar parameters of the analysed stars. Lower (l) and upper (u)
confidence levels are provided for each parameter.}\\
\hline\hline
Star & Planet & T$_{\rm eff}$ & l\_T$_{\rm eff}$ & u\_T$_{\rm eff}$ & $\log$g  & l\_$\log$g & u\_$\log$g &    v$\sin$i        & l\_v$\sin$i      & u\_v$\sin$i      \\
     &        &   (K)         & (K)              &       (K)        &  (cgs)   & (cgs)      & (cgs)      &   (kms$^{\rm -1}$) & (kms$^{\rm -1}$) & (kms$^{\rm -1}$) \\
\hline
\endfirsthead
\caption{continued.}\\
\hline\hline
Star & Planet & T$_{\rm eff}$ & l\_T$_{\rm eff}$ & u\_T$_{\rm eff}$ & $\log$g  & l\_$\log$g & u\_$\log$g &    v$\sin$i        & l\_v$\sin$i      & u\_v$\sin$i      \\
     &        &   (K)         & (K)              &       (K)        &  (cgs)   & (cgs)      & (cgs)      &   (kms$^{\rm -1}$) & (kms$^{\rm -1}$) & (kms$^{\rm -1}$) \\
\hline
\endhead
\hline
\endfoot
HD 1666 & y & 6442 & 6397 & 6487 & 4.04 & 4.02 & 4.05 & 8.35 & 8.07 & 8.68 \\
WASP-186 & y & 6600 & 6565 & 6635 & 4.27 & 4.26 & 4.28 & 16.39 & 15.86 & 17.05 \\
HAT-P-60 & y & 6858 & 6811 & 6903 & 4.09 & 4.08 & 4.11 & 10.01 & 9.35 & 11.37 \\
HIP 21152 & y & 6811 & 6785 & 6837 & 4.25 & 4.25 & 4.26 & 45.38 & 44.40 & 46.51 \\
TOI-813 & y & 6310 & 6265 & 6358 & 4.13 & 4.12 & 4.14 & 8.54 & 8.24 & 8.93 \\
EPIC 246851721 & y & 6691 & 6604 & 6791 & 4.13 & 4.10 & 4.17 & 68.21 & 56.00 & 79.96 \\
TOI-163 & y & 6597 & 6551 & 6646 & 4.21 & 4.19 & 4.22 & 15.24 & 14.81 & 15.80 \\
TOI-640 & y & 6469 & 6423 & 6516 & 4.05 & 4.04 & 4.06 & 7.95 & 7.76 & 8.19 \\
HR 2562 & y & 6697 & 6685 & 6710 & 4.30 & 4.30 & 4.31 & 44.87 & 44.41 & 45.40 \\
HD 60532 & y & 6504 & 6463 & 6549 & 3.91 & 3.90 & 3.92 & 8.22 & 7.98 & 8.51 \\
KELT-17 & y & 7360 & 7342 & 7378 & 4.16 & 4.15 & 4.16 & 43.28 & 42.77 & 43.81 \\
K2-334 & y & 6548 & 6509 & 6592 & 4.28 & 4.27 & 4.30 & 10.52 & 9.69 & 12.11 \\
MASCARA-4 & y & 7746 & 7733 & 7754 & 4.22 & 4.21 & 4.23 & 56.17 & 52.06 & 59.82 \\
TOI-1107 & y & 6698 & 6668 & 6728 & 4.16 & 4.15 & 4.16 & 15.33 & 15.05 & 15.68 \\
HATS-56 & y & 6632 & 6569 & 6697 & 4.05 & 4.03 & 4.07 & 7.67 & 7.52 & 7.88 \\
KELT-6 & y & 6664 & 6617 & 6712 & 4.26 & 4.25 & 4.27 & 7.70 & 7.56 & 7.87 \\
WASP-172 & y & 6973 & 6939 & 7011 & 4.04 & 4.03 & 4.05 & 15.60 & 15.28 & 15.97 \\
NGTS-2 & y & 6805 & 6778 & 6833 & 4.20 & 4.20 & 4.21 & 15.75 & 15.43 & 16.11 \\
WASP-189 & y & 8371 & 8356 & 8386 & 2.90 & 2.89 & 2.91 & 118.42 & 115.61 & 122.23 \\
HD 143105 & y & 6496 & 6461 & 6532 & 4.23 & 4.22 & 4.24 & 10.24 & 9.50 & 12.56 \\
TOI-2109 & y & 6821 & 6763 & 6889 & 4.20 & 4.19 & 4.22 & 108.33 & 99.30 & 117.44 \\
HAT-P-67 & y & 6589 & 6545 & 6635 & 3.92 & 3.91 & 3.93 & 38.12 & 36.16 & 40.04 \\
WASP-176 & y & 6301 & 6236 & 6369 & 4.07 & 4.05 & 4.09 & 8.02 & 7.78 & 8.37 \\
TOI-1431 & y & 7146 & 7087 & 7205 & 4.04 & 4.03 & 4.06 & 8.03 & 7.80 & 8.33 \\
TOI-1333 & y & 6738 & 6709 & 6769 & 4.17 & 4.16 & 4.18 & 15.83 & 15.44 & 16.31 \\
HD 206893 & y & 6749 & 6735 & 6765 & 4.36 & 4.36 & 4.36 & 44.64 & 44.23 & 45.07 \\
TOI-1518 & y & 7531 & 7503 & 7582 & 3.87 & 3.81 & 3.94 & 92.37 & 85.30 & 98.73 \\
HD 693 & n & 6655 & 6605 & 6704 & 4.30 & 4.28 & 4.31 & 7.57 & 7.46 & 7.71 \\
HD 7439 & n & 6818 & 6773 & 6865 & 4.26 & 4.24 & 4.27 & 7.99 & 7.80 & 8.21 \\
HD 11262 & n & 6595 & 6545 & 6645 & 4.53 & 4.52 & 4.54 & 7.65 & 7.53 & 7.82 \\
HD 13555 & n & 6770 & 6733 & 6813 & 4.13 & 4.12 & 4.14 & 8.38 & 8.14 & 8.68 \\
HD 25457 & n & 6537 & 6515 & 6561 & 4.41 & 4.41 & 4.42 & 19.08 & 18.80 & 19.42 \\
HD 30652 & n & 6738 & 6719 & 6754 & 4.37 & 4.37 & 4.38 & 19.39 & 19.07 & 19.77 \\
HD 31746 & n & 6840 & 6801 & 6877 & 4.41 & 4.40 & 4.42 & 14.06 & 13.33 & 14.65 \\
HD 33256 & n & 6832 & 6794 & 6870 & 4.22 & 4.21 & 4.23 & 9.63 & 9.14 & 10.29 \\
HD 38393 & n & 6642 & 6602 & 6681 & 4.39 & 4.38 & 4.40 & 8.73 & 8.41 & 9.15 \\
HD 40136 & n & 7143 & 7123 & 7165 & 4.23 & 4.22 & 4.23 & 19.24 & 18.92 & 19.62 \\
HD 49095 & n & 6722 & 6683 & 6763 & 4.47 & 4.46 & 4.48 & 9.02 & 8.64 & 9.50 \\
HD 49933 & n & 7142 & 7104 & 7176 & 4.38 & 4.37 & 4.38 & 10.13 & 9.55 & 12.13 \\
HD 59984 & n & 6832 & 6777 & 6883 & 4.35 & 4.34 & 4.36 & 6.40 & 6.29 & 6.56 \\
HD 68456 & n & 6847 & 6812 & 6882 & 4.07 & 4.06 & 4.08 & 13.01 & 12.69 & 13.43 \\
HD 68146 & n & 6629 & 6592 & 6666 & 4.36 & 4.35 & 4.37 & 9.38 & 8.94 & 9.98 \\
HD 76653 & n & 6670 & 6634 & 6708 & 4.38 & 4.37 & 4.39 & 12.52 & 10.43 & 12.99 \\
HD 91324 & n & 6661 & 6625 & 6697 & 4.20 & 4.19 & 4.21 & 9.30 & 8.89 & 9.79 \\
HD 91889 & n & 6500 & 6451 & 6558 & 4.32 & 4.30 & 4.33 & 7.53 & 7.42 & 7.66 \\
HD 93372 & n & 6628 & 6593 & 6663 & 4.36 & 4.35 & 4.37 & 12.85 & 12.50 & 13.32 \\
HD 94388 & n & 6566 & 6527 & 6606 & 4.08 & 4.06 & 4.09 & 9.27 & 8.85 & 9.82 \\
HD 100563 & n & 6684 & 6656 & 6712 & 4.34 & 4.33 & 4.34 & 15.16 & 14.93 & 15.44 \\
HD 101198 & n & 6549 & 6495 & 6617 & 4.23 & 4.20 & 4.26 & 7.50 & 7.39 & 7.62 \\
HD 104731 & n & 6894 & 6868 & 6918 & 4.26 & 4.26 & 4.27 & 15.65 & 15.37 & 15.99 \\
HD 114642 & n & 6648 & 6618 & 6679 & 3.97 & 3.96 & 3.98 & 14.95 & 14.76 & 15.17 \\
HD 124850 & n & 6504 & 6485 & 6526 & 3.92 & 3.91 & 3.92 & 16.11 & 15.75 & 16.52 \\
HD 125276 & n & 6948 & 6895 & 7018 & 4.66 & 4.65 & 4.67 & 6.37 & 6.25 & 6.53 \\
HD 128167 & n & 7235 & 7201 & 7273 & 4.42 & 4.41 & 4.43 & 8.79 & 8.50 & 9.15 \\
HD 128020 & n & 6580 & 6531 & 6630 & 4.37 & 4.36 & 4.38 & 7.78 & 7.63 & 7.98 \\
HD 128898 & n & 7615 & 7574 & 7656 & 4.13 & 4.13 & 4.14 & 15.99 & 15.45 & 16.69 \\
HD 133469 & n & 6684 & 6665 & 6703 & 4.38 & 4.38 & 4.39 & 25.30 & 24.85 & 25.83 \\
HD 138763 & n & 6396 & 6357 & 6436 & 4.51 & 4.50 & 4.52 & 8.90 & 8.54 & 9.37 \\
HD 139211 & n & 6576 & 6526 & 6626 & 4.28 & 4.27 & 4.30 & 7.88 & 7.71 & 8.10 \\
HD 184985 & n & 6503 & 6453 & 6561 & 4.09 & 4.08 & 4.11 & 7.54 & 7.44 & 7.69 \\
HD 191862 & n & 6733 & 6699 & 6769 & 4.34 & 4.33 & 4.35 & 9.30 & 8.90 & 9.82 \\
HD 196385 & n & 7268 & 7244 & 7294 & 4.30 & 4.29 & 4.30 & 15.85 & 15.54 & 16.19 \\
HD 198390 & n & 6896 & 6858 & 6934 & 4.38 & 4.38 & 4.39 & 8.77 & 8.47 & 9.14 \\
HD 199260 & n & 6616 & 6588 & 6643 & 4.46 & 4.45 & 4.46 & 15.28 & 15.04 & 15.58 \\
HD 210302 & n & 6612 & 6582 & 6643 & 4.31 & 4.30 & 4.32 & 15.12 & 14.89 & 15.40 \\
HD 211976 & n & 6750 & 6707 & 6796 & 4.38 & 4.37 & 4.39 & 7.92 & 7.75 & 8.15 \\
HD 213845 & n & 6759 & 6750 & 6773 & 4.30 & 4.29 & 4.30 & 35.12 & 34.72 & 35.59 \\
HD 219482 & n & 6561 & 6519 & 6606 & 4.44 & 4.43 & 4.45 & 8.62 & 8.31 & 9.02 \\
HD 220729 & n & 6916 & 6896 & 6937 & 4.23 & 4.22 & 4.23 & 22.19 & 21.75 & 22.49 \\
HD 222368 & n & 6532 & 6488 & 6584 & 4.25 & 4.23 & 4.26 & 7.88 & 7.70 & 8.11 \\
\end{longtable}

\onecolumn
{\scriptsize
\begin{longtable}{lccccccccccc}
\caption{\label{abd_param_table}
 Derived abundances, [X/H] (dex).}\\
\hline\hline
Star & Planet & C & O & Mg & Si &  Ca &  Ti & Fe &  Ni & Ba \\
\hline
\endfirsthead
\caption{continued.}\\
\hline\hline
Star & Planet & C & O & Mg & Si &  Ca &  Ti  Fe &  Ni & Ba \\
\hline
\endhead
\hline
\endfoot
HD 1666 & y & 0.28$\pm$0.14 & -0.03$\pm$0.07 & -0.04$\pm$0.07 & -- & -0.01$\pm$0.02 & 0.02$\pm$0.12 & 0.11$\pm$0.02 & 0.04$\pm$0.06 & 0.02$\pm$0.06 \\
WASP-186 & y & 0.11$\pm$0.04 & -0.21$\pm$0.11 & -0.12$\pm$0.07 & -0.11$\pm$0.22 & -0.13$\pm$0.03 & -0.03$\pm$0.12 & -0.01$\pm$0.02 & -0.10$\pm$0.06 & -0.18$\pm$0.06 \\
HAT-P-60 & y & -0.08$\pm$0.03 & -0.13$\pm$0.07 & -0.28$\pm$0.07 & -0.37$\pm$0.04 & -0.27$\pm$0.02 & -0.17$\pm$0.13 & -0.22$\pm$0.05 & -0.25$\pm$0.06 & -0.19$\pm$0.06 \\
HIP 21152 & y & 0.08$\pm$0.09 & 0.09$\pm$0.06 & -0.09$\pm$0.07 & -- & -0.07$\pm$0.15 & -- & -0.15$\pm$0.07 & 0.16$\pm$0.06 & -0.22$\pm$0.06 \\
TOI-813 & y & 0.14$\pm$0.04 & -0.13$\pm$0.06 & -0.05$\pm$0.07 & -- & 0.02$\pm$0.02 & -0.03$\pm$0.12 & 0.08$\pm$0.04 & -0.02$\pm$0.06 & 0.04$\pm$0.06 \\
EPIC 246851721 & y & 0.70$\pm$0.15 & 0.81$\pm$0.09 & -0.48$\pm$0.07 & -- & 0.06$\pm$0.07 & -- & -0.34$\pm$0.17 & 0.18$\pm$0.09 & -0.46$\pm$0.23 \\
TOI-163 & y & 0.14$\pm$0.05 & 0.34$\pm$0.08 & -0.09$\pm$0.07 & -- & 0.05$\pm$0.04 & -0.05$\pm$0.19 & 0.02$\pm$0.07 & -0.18$\pm$0.06 & -0.07$\pm$0.06 \\
TOI-640 & y & 0.05$\pm$0.03 & -0.22$\pm$0.06 & -0.12$\pm$0.07 & -0.17$\pm$0.23 & -0.08$\pm$0.05 & -0.03$\pm$0.13 & 0.00$\pm$0.01 & -0.06$\pm$0.06 & 0.06$\pm$0.06 \\
HR 2562 & y & -0.00$\pm$0.16 & 0.01$\pm$0.10 & -0.10$\pm$0.07 & -- & -0.09$\pm$0.11 & -- & -0.11$\pm$0.07 & 0.07$\pm$0.06 & -0.18$\pm$0.08 \\
HD 60532 & y & -0.02$\pm$0.05 & -0.29$\pm$0.06 & -0.22$\pm$0.07 & -0.30$\pm$0.16 & -0.17$\pm$0.03 & -0.08$\pm$0.13 & -0.09$\pm$0.02 & -0.18$\pm$0.06 & 0.04$\pm$0.06 \\
KELT-17 & y & -0.70$\pm$0.17 & -0.66$\pm$0.06 & -0.40$\pm$0.07 & -- & -- & -0.00$\pm$0.24 & 0.17$\pm$0.11 & 0.38$\pm$0.06 & 1.07$\pm$0.06 \\
K2-334 & y & 0.11$\pm$0.03 & -0.11$\pm$0.06 & -0.09$\pm$0.07 & -- & -0.05$\pm$0.06 & -0.03$\pm$0.11 & 0.03$\pm$0.02 & -0.04$\pm$0.06 & -0.15$\pm$0.06 \\
MASCARA-4 & y & -0.04$\pm$0.06 & -0.09$\pm$0.06 & -0.65$\pm$0.07 & -- & -0.57$\pm$0.08 & -- & -0.22$\pm$0.12 & 0.03$\pm$0.06 & 0.02$\pm$0.06 \\
TOI-1107 & y & 0.09$\pm$0.06 & -0.02$\pm$0.06 & -0.24$\pm$0.07 & -- & -0.19$\pm$0.07 & -0.12$\pm$0.14 & -0.10$\pm$0.02 & -0.25$\pm$0.06 & -0.11$\pm$0.06 \\
HATS-56 & y & 0.21$\pm$0.08 & -0.07$\pm$0.06 & -0.20$\pm$0.07 & -- & -0.12$\pm$0.08 & -0.05$\pm$0.12 & -0.01$\pm$0.04 & -0.05$\pm$0.06 & -0.01$\pm$0.06 \\
KELT-6 & y & -0.09$\pm$0.06 & -0.38$\pm$0.06 & -0.23$\pm$0.07 & -0.39$\pm$0.22 & -0.29$\pm$0.07 & -0.13$\pm$0.15 & -0.21$\pm$0.05 & -0.21$\pm$0.06 & -0.30$\pm$0.06 \\
WASP-172 & y & 0.13$\pm$0.08 & -0.04$\pm$0.06 & -0.22$\pm$0.07 & -- & -0.15$\pm$0.11 & -0.00$\pm$0.11 & -0.17$\pm$0.03 & -0.28$\pm$0.06 & -0.01$\pm$0.06 \\
NGTS-2 & y & 0.08$\pm$0.05 & -0.09$\pm$0.06 & -0.21$\pm$0.07 & -0.25$\pm$0.23 & -0.20$\pm$0.09 & -0.11$\pm$0.15 & -0.11$\pm$0.04 & -0.20$\pm$0.06 & -0.21$\pm$0.06 \\
WASP-189 & y & -- & -0.39$\pm$0.06 & 0.28$\pm$0.07 & -0.06$\pm$0.05 & -- & -- & 0.00$\pm$0.21 & 0.79$\pm$0.06 & 0.74$\pm$0.06 \\
HD 143105 & y & 0.09$\pm$0.08 & -0.11$\pm$0.08 & -0.10$\pm$0.07 & -0.04$\pm$0.11 & -0.05$\pm$0.04 & -0.06$\pm$0.12 & 0.02$\pm$0.01 & -0.06$\pm$0.06 & -0.03$\pm$0.06 \\
TOI-2109 & y & -- & -- & -0.01$\pm$0.07 & -- & -0.96$\pm$0.10 & -- & -- & 0.43$\pm$0.08 & -1.26$\pm$0.20 \\
HAT-P-67 & y & 0.01$\pm$0.06 & -- & -0.17$\pm$0.07 & -- & -0.42$\pm$0.09 & -0.06$\pm$0.07 & -0.15$\pm$0.07 & -0.15$\pm$0.08 & -0.28$\pm$0.10 \\
WASP-176 & y & 0.14$\pm$0.06 & -- & -0.08$\pm$0.07 & 0.03$\pm$0.23 & -0.16$\pm$0.10 & -0.01$\pm$0.16 & 0.08$\pm$0.02 & -0.07$\pm$0.06 & -0.21$\pm$0.14 \\
TOI-1431 & y & -0.27$\pm$0.08 & -0.39$\pm$0.06 & -0.42$\pm$0.07 & -- & -- & -0.08$\pm$0.15 & 0.05$\pm$0.08 & 0.13$\pm$0.06 & 0.87$\pm$0.06 \\
TOI-1333 & y & 0.18$\pm$0.16 & -0.07$\pm$0.06 & -0.18$\pm$0.07 & -- & -0.15$\pm$0.06 & -0.04$\pm$0.15 & -0.05$\pm$0.03 & -0.12$\pm$0.06 & -0.19$\pm$0.06 \\
HD 206893 & y & -- & -0.10$\pm$0.06 & -0.38$\pm$0.07 & -0.31$\pm$0.24 & -0.25$\pm$0.01 & -0.12$\pm$0.24 & -0.26$\pm$0.04 & -0.10$\pm$0.06 & -0.20$\pm$0.06 \\
TOI-1518 & y & -0.15$\pm$0.11 & -0.04$\pm$0.06 & -0.12$\pm$0.07 & -- & -- & -- & -0.21$\pm$0.04 & 0.31$\pm$0.07 & -0.03$\pm$0.10 \\
HD 693 & n & -0.17$\pm$0.06 & -0.50$\pm$0.06 & -0.23$\pm$0.07 & -0.46$\pm$0.21 & -0.34$\pm$0.09 & -0.14$\pm$0.14 & -0.24$\pm$0.06 & -0.24$\pm$0.06 & -0.39$\pm$0.06 \\
HD 7439 & n & -0.11$\pm$0.06 & -0.37$\pm$0.06 & -0.28$\pm$0.07 & -0.44$\pm$0.17 & -0.33$\pm$0.07 & -0.15$\pm$0.15 & -0.24$\pm$0.07 & -0.26$\pm$0.06 & -0.31$\pm$0.06 \\
HD 11262 & n & -0.11$\pm$0.05 & -0.42$\pm$0.06 & -0.22$\pm$0.07 & -0.38$\pm$0.23 & -0.25$\pm$0.04 & -0.14$\pm$0.16 & -0.14$\pm$0.03 & -0.22$\pm$0.06 & -0.19$\pm$0.06 \\
HD 13555 & n & -0.05$\pm$0.05 & -0.31$\pm$0.06 & -0.28$\pm$0.07 & -0.39$\pm$0.20 & -0.29$\pm$0.06 & -0.14$\pm$0.15 & -0.18$\pm$0.04 & -0.23$\pm$0.06 & -0.20$\pm$0.06 \\
HD 25457 & n & 0.03$\pm$0.04 & -0.13$\pm$0.06 & -0.16$\pm$0.07 & -- & -0.03$\pm$0.02 & -0.09$\pm$0.15 & -0.01$\pm$0.04 & -0.19$\pm$0.06 & 0.10$\pm$0.06 \\
HD 30652 & n & 0.08$\pm$0.04 & -0.09$\pm$0.06 & -0.20$\pm$0.07 & -- & -0.15$\pm$0.06 & -0.12$\pm$0.18 & -0.06$\pm$0.02 & -0.19$\pm$0.06 & -0.21$\pm$0.06 \\
HD 31746 & n & 0.02$\pm$0.05 & -0.21$\pm$0.06 & -0.27$\pm$0.07 & -0.28$\pm$0.23 & -0.23$\pm$0.07 & -0.14$\pm$0.16 & -0.13$\pm$0.06 & -0.23$\pm$0.06 & -0.21$\pm$0.06 \\
HD 33256 & n & -0.14$\pm$0.05 & -0.37$\pm$0.06 & -0.24$\pm$0.07 & -0.42$\pm$0.13 & -0.30$\pm$0.08 & -0.16$\pm$0.14 & -0.23$\pm$0.07 & -0.28$\pm$0.06 & -0.28$\pm$0.06 \\
HD 38393 & n & 0.02$\pm$0.05 & -0.25$\pm$0.06 & -0.20$\pm$0.07 & -0.25$\pm$0.23 & -0.18$\pm$0.04 & -0.10$\pm$0.16 & -0.08$\pm$0.03 & -0.14$\pm$0.06 & -0.20$\pm$0.06 \\
HD 40136 & n & 0.13$\pm$0.06 & -0.08$\pm$0.06 & -0.21$\pm$0.07 & -- & -0.26$\pm$0.14 & -0.08$\pm$0.12 & -0.16$\pm$0.06 & -0.22$\pm$0.06 & -0.15$\pm$0.06 \\
HD 49095 & n & -0.12$\pm$0.06 & -0.41$\pm$0.06 & -0.22$\pm$0.07 & -0.41$\pm$0.18 & -0.27$\pm$0.05 & -0.14$\pm$0.17 & -0.17$\pm$0.05 & -0.24$\pm$0.06 & -0.31$\pm$0.06 \\
HD 49933 & n & -0.22$\pm$0.04 & -0.55$\pm$0.06 & -0.39$\pm$0.07 & -0.58$\pm$0.09 & -0.46$\pm$0.15 & -0.21$\pm$0.14 & -0.33$\pm$0.11 & -0.33$\pm$0.06 & -0.38$\pm$0.06 \\
HD 59984 & n & -0.47$\pm$0.06 & -0.76$\pm$0.10 & -0.14$\pm$0.07 & -0.72$\pm$0.08 & -0.40$\pm$0.04 & -0.18$\pm$0.20 & -0.37$\pm$0.07 & -0.37$\pm$0.06 & -0.52$\pm$0.06 \\
HD 68456 & n & -0.03$\pm$0.06 & -0.23$\pm$0.06 & -0.03$\pm$0.07 & -0.38$\pm$0.16 & -0.28$\pm$0.11 & -0.14$\pm$0.12 & -0.24$\pm$0.09 & -0.26$\pm$0.06 & -0.51$\pm$0.06 \\
HD 68146 & n & 0.05$\pm$0.03 & -0.17$\pm$0.06 & -0.18$\pm$0.07 & -0.18$\pm$0.21 & -0.15$\pm$0.02 & -0.09$\pm$0.16 & -0.05$\pm$0.02 & -0.09$\pm$0.06 & -0.20$\pm$0.06 \\
HD 76653 & n & 0.06$\pm$0.04 & -0.14$\pm$0.06 & -0.15$\pm$0.07 & -0.17$\pm$0.22 & -0.10$\pm$0.03 & -0.09$\pm$0.16 & -0.02$\pm$0.02 & -0.11$\pm$0.06 & -0.06$\pm$0.06 \\
HD 91324 & n & -0.10$\pm$0.03 & -0.45$\pm$0.06 & -0.23$\pm$0.07 & -0.44$\pm$0.17 & -0.31$\pm$0.06 & -0.15$\pm$0.16 & -0.20$\pm$0.04 & -0.28$\pm$0.06 & -0.31$\pm$0.06 \\
HD 91889 & n & -0.10$\pm$0.07 & -0.41$\pm$0.06 & -0.21$\pm$0.07 & -- & -0.24$\pm$0.03 & -0.12$\pm$0.17 & -0.13$\pm$0.05 & -0.10$\pm$0.06 & -0.24$\pm$0.06 \\
HD 93372 & n & 0.10$\pm$0.05 & -0.11$\pm$0.06 & -0.16$\pm$0.07 & -0.14$\pm$0.25 & -0.10$\pm$0.03 & -0.08$\pm$0.15 & -0.01$\pm$0.02 & -0.10$\pm$0.06 & -0.14$\pm$0.06 \\
HD 94388 & n & 0.10$\pm$0.05 & -0.12$\pm$0.06 & -0.11$\pm$0.07 & -0.14$\pm$0.20 & -0.07$\pm$0.03 & -0.04$\pm$0.14 & 0.01$\pm$0.02 & -0.09$\pm$0.06 & -0.01$\pm$0.06 \\
HD 100563 & n & 0.15$\pm$0.06 & -0.02$\pm$0.06 & -0.16$\pm$0.07 & -- & -0.10$\pm$0.05 & -0.09$\pm$0.16 & -0.01$\pm$0.02 & -0.12$\pm$0.06 & -0.19$\pm$0.06 \\
HD 101198 & n & -0.04$\pm$0.06 & -0.40$\pm$0.06 & -0.20$\pm$0.07 & -- & -0.26$\pm$0.07 & -0.11$\pm$0.15 & -0.14$\pm$0.05 & -0.10$\pm$0.06 & -0.25$\pm$0.06 \\
HD 104731 & n & 0.04$\pm$0.05 & -0.17$\pm$0.06 & -0.21$\pm$0.07 & -0.31$\pm$0.22 & -0.25$\pm$0.13 & -0.12$\pm$0.12 & -0.17$\pm$0.06 & -0.24$\pm$0.06 & -0.29$\pm$0.06 \\
HD 114642 & n & 0.00$\pm$0.05 & -0.25$\pm$0.06 & -0.20$\pm$0.07 & -0.34$\pm$0.20 & -0.24$\pm$0.07 & -0.14$\pm$0.14 & -0.14$\pm$0.03 & -0.27$\pm$0.06 & -0.07$\pm$0.06 \\
HD 124850 & n & 0.03$\pm$0.04 & -0.23$\pm$0.06 & -0.18$\pm$0.07 & -0.26$\pm$0.22 & -0.14$\pm$0.06 & -0.10$\pm$0.14 & -0.09$\pm$0.02 & -0.25$\pm$0.06 & 0.25$\pm$0.06 \\
HD 125276 & n & -0.46$\pm$0.12 & -0.79$\pm$0.11 & -0.13$\pm$0.07 & -0.61$\pm$0.08 & -0.40$\pm$0.05 & -0.20$\pm$0.20 & -0.35$\pm$0.09 & -0.26$\pm$0.06 & -0.41$\pm$0.06 \\
HD 128167 & n & -0.20$\pm$0.04 & -0.53$\pm$0.06 & -0.35$\pm$0.07 & -0.57$\pm$0.10 & -0.42$\pm$0.20 & -0.18$\pm$0.10 & -0.29$\pm$0.13 & -0.28$\pm$0.06 & -0.34$\pm$0.06 \\
HD 128020 & n & -0.06$\pm$0.05 & -0.39$\pm$0.06 & -0.23$\pm$0.07 & -0.35$\pm$0.22 & -0.25$\pm$0.04 & -0.13$\pm$0.16 & -0.13$\pm$0.02 & -0.18$\pm$0.06 & -0.28$\pm$0.06 \\
HD 128898 & n & -0.35$\pm$0.03 & -0.24$\pm$0.06 & -0.29$\pm$0.07 & 0.00$\pm$0.20 & -0.11$\pm$0.19 & -0.21$\pm$0.06 & -0.21$\pm$0.17 & -0.62$\pm$0.09 & -0.72$\pm$0.06 \\
HD 133469 & n & -0.03$\pm$0.03 & -0.11$\pm$0.06 & -0.15$\pm$0.07 & -0.14$\pm$0.21 & -0.09$\pm$0.01 & -0.06$\pm$0.16 & -0.05$\pm$0.02 & -0.13$\pm$0.06 & -0.04$\pm$0.06 \\
HD 138763 & n & -0.01$\pm$0.04 & -0.34$\pm$0.08 & -0.13$\pm$0.07 & -0.19$\pm$0.24 & -0.01$\pm$0.04 & -0.08$\pm$0.11 & 0.02$\pm$0.05 & -0.09$\pm$0.06 & 0.24$\pm$0.06 \\
HD 139211 & n & 0.05$\pm$0.05 & -0.19$\pm$0.06 & -0.17$\pm$0.07 & -- & -0.14$\pm$0.04 & -0.07$\pm$0.14 & -0.04$\pm$0.01 & -0.09$\pm$0.06 & -0.14$\pm$0.06 \\
HD 184985 & n & 0.11$\pm$0.05 & -0.18$\pm$0.06 & -0.15$\pm$0.07 & -0.21$\pm$0.23 & -0.14$\pm$0.05 & -0.09$\pm$0.14 & -0.04$\pm$0.02 & -0.12$\pm$0.06 & -0.07$\pm$0.06 \\
HD 191862 & n & -0.08$\pm$0.03 & -0.37$\pm$0.06 & -0.25$\pm$0.07 & -0.38$\pm$0.16 & -0.26$\pm$0.06 & -0.13$\pm$0.16 & -0.16$\pm$0.04 & -0.23$\pm$0.06 & -0.30$\pm$0.06 \\
HD 196385 & n & 0.01$\pm$0.06 & -0.26$\pm$0.06 & -0.28$\pm$0.07 & -0.40$\pm$0.23 & -0.35$\pm$0.17 & -0.15$\pm$0.11 & -0.22$\pm$0.08 & -0.25$\pm$0.06 & -0.21$\pm$0.06 \\
HD 198390 & n & -0.11$\pm$0.04 & -0.40$\pm$0.06 & -0.31$\pm$0.07 & -0.46$\pm$0.15 & -0.34$\pm$0.09 & -0.16$\pm$0.15 & -0.23$\pm$0.07 & -0.26$\pm$0.06 & -0.33$\pm$0.06 \\
HD 199260 & n & -0.01$\pm$0.04 & -0.27$\pm$0.06 & -0.21$\pm$0.07 & -0.27$\pm$0.23 & -0.16$\pm$0.04 & -0.14$\pm$0.16 & -0.08$\pm$0.02 & -0.23$\pm$0.06 & -0.05$\pm$0.06 \\
HD 210302 & n & 0.11$\pm$0.05 & -0.08$\pm$0.06 & -0.17$\pm$0.07 & -0.13$\pm$0.24 & -0.08$\pm$0.04 & -0.08$\pm$0.15 & -0.00$\pm$0.01 & -0.11$\pm$0.06 & -0.14$\pm$0.06 \\
HD 211976 & n & -0.02$\pm$0.06 & -0.29$\pm$0.06 & -0.26$\pm$0.07 & -0.33$\pm$0.24 & -0.24$\pm$0.06 & -0.11$\pm$0.16 & -0.13$\pm$0.05 & -0.16$\pm$0.06 & -0.24$\pm$0.06 \\
HD 213845 & n & 0.06$\pm$0.05 & -0.03$\pm$0.06 & -0.13$\pm$0.07 & -- & -0.08$\pm$0.06 & -0.04$\pm$0.17 & -0.09$\pm$0.02 & 0.01$\pm$0.06 & -0.17$\pm$0.06 \\
HD 219482 & n & 0.01$\pm$0.05 & -0.23$\pm$0.06 & -0.17$\pm$0.07 & -0.20$\pm$0.24 & -0.11$\pm$0.01 & -0.10$\pm$0.15 & -0.03$\pm$0.01 & -0.12$\pm$0.06 & 0.04$\pm$0.06 \\
HD 220729 & n & 0.12$\pm$0.03 & -0.01$\pm$0.06 & -0.23$\pm$0.07 & -- & -0.18$\pm$0.06 & -0.08$\pm$0.16 & -0.10$\pm$0.05 & -0.18$\pm$0.06 & -0.18$\pm$0.06 \\
HD 222368 & n & -0.01$\pm$0.05 & -0.29$\pm$0.06 & -0.17$\pm$0.07 & -0.29$\pm$0.22 & -0.19$\pm$0.05 & -0.10$\pm$0.15 & -0.09$\pm$0.02 & -0.18$\pm$0.06 & -0.14$\pm$0.06 \\
\end{longtable}
}

\end{appendix}

\end{document}